\documentclass[prd,twocolumn,floatfix,amsmath,nofootinbib,amssymb,floatfix]{revtex4}
\usepackage{graphicx,color,dcolumn,booktabs,bm}
\usepackage{longtable,lscape}
\usepackage{pdfpages}
\usepackage{txfonts}
\usepackage{overpic}
\usepackage{amssymb}
\usepackage{makecell}
\usepackage{indentfirst}
\usepackage{feynmf}   
\usepackage{slashed}  
\usepackage{cases}
\usepackage{color}
\usepackage{multirow}
\usepackage{threeparttable}
\usepackage{epstopdf}
\usepackage{enumerate}
\usepackage{subfigure}
\usepackage{diagbox}
\usepackage{graphicx,color,dcolumn,booktabs,bm}
\usepackage{mathrsfs}
\usepackage{cancel}
\usepackage{float}
\usepackage[misc]{ifsym}
\usepackage[colorlinks,
            citecolor=blue,
            anchorcolor=red,
            menucolor=red,
            linkcolor=red,
            filecolor=red,
            runcolor=red,
            urlcolor=blue,
            frenchlinks=red]{hyperref}
\usepackage{array}
\usepackage{booktabs}

\begin{document}
\title{Systematic analysis of the transition form factors of $B_c$ to $D$-wave charmonia and corresponding semileptonic decays}
\author{Jie Lu$^{1,2}$}
\email{l17693567997@163.com}
\author{Dian-Yong Chen$^{1,3}$}
\email{chendy@seu.edu.cn}
\author{Guo-Liang Yu$^{2}$}
\email{yuguoliang2011@163.com}
\author{Zhi-Gang Wang$^{2}$}
\email{zgwang@aliyun.com}

\affiliation{$^1$ School of Physics, Southeast University, Nanjing 210094, People's Republic of
China\\$^2$ Department of Mathematics and Physics, North China
Electric Power University, Baoding 071003, People's Republic of
China\\$^3$ Lanzhou Center for Theoretical Physics, Lanzhou University, Lanzhou 730000, People's Republic of
China}
\date{\today}

\begin{abstract}
In this article, the vector, axial vector and tensor form factors of $B_c$ to $D$-wave charmonia $\psi_J$ ($J=1,2,3$) and $\eta_{c2}$ are analyzed within the framework of three-point QCD sum rules. With the predicted vector and axial vector form factors, we study the decay widths and branching ratios of semileptonic decays $B_c\to\psi_Jl\bar{\nu_l}$ and $B_c\to\eta_{c2}l\bar{\nu_l}$. Our results indicate that the semileptonic decay branching ratios of $B_c$ to $D$-wave charmonia decreases as the total angular momentum of the final state $D$-wave charmonia increases. For decay processes $B_c\to \psi_1l\bar{\nu}_l$, $B_c\to \psi_2[\eta_{c2}]l\bar{\nu}_l$ and $B_c\to \psi_3l\bar{\nu}_l$, the branching ratios can reach the order of magnitude $10^{-3}$, $10^{-4}$ and $10^{-5}$, respectively, which can provide a valuable reference for future experimental measurements for $B_c$ meson by LHCb collaboration. Furthermore, these results also can provide useful information to study the properties of $B_c$ meson and $D$-wave charmonia.
\end{abstract}


\maketitle

\section{Introduction}\label{sec1}
The decay processes of the $B_c$ meson have attracted considerable interest due to its distinctive composition in the Standard Model (SM). As the only quarkonium with mixed heavy flavor quarks, $B_c$ meson can not annihilate into gluon or photon via strong or electromagnetic interactions, hence its decays can only occur via weak interaction, which includes the $b\to c(u)$, $c\to s(d)$ transitions and the weak annihilation. Since each heavy flavor quark in $B_c$ meson can decay individually and the other one acts as a spectator, the $B_c$ meson theoretically has rich decay channels, which provides an ideal platform to test various theoretical approaches such as the perturbation QCD~\cite{Rui:2016opu,Rui:2017pre,Duan:2017rhr,Liu:2023kxr,Liu:2025ipe,Zhao:2025zht}, the QCD factorization~\cite{Hsiao:2016pml,Chang:2018mva}, the QCD sum rules~\cite{Colangelo:1992cx,Kiselev:2000pp,Azizi:2009ny,Wu:2024gcq,Wu:2024dzn,Lu:2025bvi}, the light-cone QCD sum rules~\cite{Wang:2007fs,Leljak:2019eyw,Bordone:2022drp}, various quark models~\cite{Ivanov:2006ni,Faustov:2022ybm,Zhang:2023ypl,Sun:2023iis} and others~\cite{Chang:1992pt,Qiao:2012hp}. On the experimental aspect, since the discovery of $B_c$ meson by CDF collaboration via the decay process $B_c\to J/\psi l\bar{\nu}_l$ in 1.8 TeV $p\bar{p}$ collisions~\cite{CDF:1998axz}, LHCb collaboration has observed many $B_c$ decay channels such as $B^+_c\to J/\psi\pi^+\pi^-[K^+,K^-]\pi^+$, $B^+_c\to J/\psi\pi^+[K^+,D_s^{(*)+}]$, $B_c^+\to B^0_s\pi^+$ and $B_c^+\to\chi_{c0}\pi^+$~\cite{LHCb:2012ag,LHCb:2012ihf,LHCb:2013hwj,LHCb:2013kwl,LHCb:2013rud,LHCb:2013xlg,LHCb:2015azd,LHCb:2017ogk,LHCb:2023fqn}. In addition, it was estimated that the inclusive production cross sections of $B_c$ meson and its resonance states at the LHCb collaboration is at level of 1 $\mu$b for $\sqrt{s}= \sqrt{14}$ TeV. This means around $O$(10$^9$) $B_c$ meson can be anticipated in 1 fb$^{-1}$~\cite{Gao:2010zzc}, which will provide sufficient events to study the decay processes of $B_c$ meson.

The weak decays of the $B_c$ meson mainly include semileptonic and nonleptonic processes. Of particular interest are the semileptonic decays, which offer a cleaner theoretical framework due to the absence of strong interaction complications in the final-state. Taking the $B_c^-$ meson as an example, at the quark level, these decays proceed via the weak decay of either the $b$ or $\bar{c}$ quark, while the other acts as a spectator. Specifically, the $b$ quark decay follows $b\to c(u) l^-\bar{\nu}_l$ and the $\bar{c}$ quark decay follows $\bar{c}\to \bar{s}(\bar{d})l^-\bar{\nu}_l$ ($l=e,\mu,\tau$). The quark originating from the weak decay then hadronizes with the spectator quark to form the final-state meson. In this work, we focus exclusively on semileptonic $B_c$ decays in which the final hadronic state is a charmonium.

The charmonium is made of a charmed quark and its anti-quark, which can provide a suitable place to study QCD in the non-perturbative region. Since the observation of the $J/\psi$ in 1974~\cite{E598:1974sol}, many charmonium states, such as the $\eta_c$, $\eta_c'$, $h_c$, $\chi_{c0}$, $\chi_{c1}$ and $\chi_{c2}$ were reported by the experimental collaborations~\cite{Tanenbaum:1975ef,Whitaker:1976hb,DASP:1978dns,Partridge:1980vk,Edwards:1981ex}. The $D$-wave charmonium has the orbital angular momentum $L=2$. According to the $L-S$ coupling in quark model, there are four ground state $D$-wave charmonia, which are the spin-triplet state $\psi_{1,2,3}$ ($^{2s+1}L_J=^{3}D_{1,2,3}$) and spin-singlet state $\eta_{c2}$ ($^{2s+1}L_J=^{1}D_2$), respectively. The ground state $^3D_1$ charmonium candidate was firstly observed with the mass being $3772\pm6$ MeV in 1977~\cite{Rapidis:1977cv}. For ground state $^3D_2$ charmonium, it has been detected several times in experiments~\cite{Belle:2013ewt,BESIII:2015iqd,LHCb:2020fvo,BESIII:2022yga,BESIII:2022cyq}. A new narrow charmonium state named as $X(3842)$ was observed by the LHCb collaboration in decay model $X(3824)\to D^0\bar{D}^0/D^+D^-$ with very high statistical significance~\cite{LHCb:2019lnr}, which can be considered as a candidate for ground state $^3D_3$ charmonium. The ground state $^1D_2$ charmonium has not been determined yet experimentally. On theoretical aspect, some phenomenological approaches were employed to study the mass spectrum of $D$-wave charmonia, such as the QCD sum rules~\cite{Xin:2024xbp}, the coupling channel model~\cite{Man:2024mvl} and various quark models~\cite{Deng:2016stx,Bhavsar:2018umj,Kher:2018wtv,Wang:2019mhs}. 

The semileptonic decay processes of $B_c$ to charmonia are all dominated by the decay process $b \to cl\bar{\nu}_l$ at quark level, and can be described using the electroweak effective Hamiltonian in the SM. However, the matrix elements involving hadronic transition can not be obtained through perturbative calculations. Generally, we define them as various form factors based on conservation of quantum numbers and Lorentz invariance. Then, these form factors can be obtained by some non-perturbative approaches. As one of the most powerful non-perturbative approach, the QCD sum rules are widely used in studying the properties of hadrons~\cite{Shifman:1978bx,Shifman:1978by,Colangelo:2000dp,Wang:2025sic}. In our previous works, the form factors of $B_c$ to $S$ and $P$-wave charmonia were systematically analyzed by three-point QCD sum rules, and the corresponding semileptonic and nonleptonic decay processes were studied~\cite{Wu:2024gcq,Lu:2025bvi}. However, until now, there has been no work on the study of the form factors and corresponding decays of $B_c$ to $D$-wave charmonia in the QCD sum rules. Moreover, in recent work, the authors constructed the interpolate currents of $D$-wave charmonia and studied their masses and decay constants in two-point QCD sum rules, which provide important materials for the studying of their decay properties~\cite{Xin:2024xbp}. As a continuation and expansion of our previous work, we systematically analyze the vector, axial vector and tensor form factors of $B_c$ to $D$-wave charmonia $\psi_J$ and $\eta_{c2}$ by using three-point QCD sum rules in the present work. With the estimated vector and axial vector form factors, we also study the semileptonic decay processes $B_c\to\psi_Jl\bar{\nu}_l$ and $B_c\to\eta_{c2}l\bar{\nu}_l$.

This paper is organized as follows. After introduction in Sec.~\ref{sec1}, the vector, axial vector and tensor form factors of $B_c$ to $D$-wave charmonia are analyzed by the three-point QCD sum rules in Sec.~\ref{sec2}. With the estimated vector and axial vector form factors, we analyze the corresponding semileptonic decay processes in Sec.~\ref{sec3}. Sec.~\ref{sec4} is employed to present the numerical results and discussions and Sec.~\ref{sec5} is conclusion part. Some complicated formulas are shown in Appendix~\ref{Sec:AppA}.
\section{Three-point QCD sum rules for transition form factors}\label{sec2}
To study the form factors of $B_{c}$ to $D$-wave charmonia, we firstly construct the following three-point correlation function:
\begin{eqnarray}\label{eq:1}
\notag
\Pi (p,p') &&= i^2\int d^4xd^4ye^{ip'\cdot z}e^{i(p-p')\cdot y}e^{-ip\cdot x} \\
&& \times \left\langle 0 \right| \mathcal{T}\{ J_X(z)\tilde J(y)J_{B_c}^\dagger(x)\} \left| 0 \right\rangle|_{z\to0},
\end{eqnarray}
where $\mathcal{T}$ represents the time ordered product, $J_{B_c}$ and $J_{X}$ ($X=\psi_J$ and $\eta_{c2}$, $J=1,2,3$) denote the interpolating currents of $B_{c}$ meson and $D$-wave charmonia, respectively.  $\tilde{J}$ is the transition currents. These currents are taken as the following forms~\cite{Xin:2024xbp,Wang:2024fwc}:
\begin{eqnarray}\label{eq:2}
\notag
J_{B_c}(x) &&= \bar c(x)i\gamma _5b(x),\\
\notag
\tilde J(y) &&= \bar c(y)\Gamma b(y),\\
\notag
J^{\psi _1}_\alpha(z) &&= \bar c(z)\stackrel{\leftrightarrow}{D}_{\alpha'} \stackrel{\leftrightarrow}{D}_{\beta'}\gamma_{\rho} \left(g^{\alpha'\beta'} g^{\rho\alpha}+g^{\alpha'\rho} g^{\beta'\alpha}+g^{\rho\beta'} g^{\alpha'\alpha} \right)c(z),\\
\notag
J^{\psi _2}_{\alpha\beta}(z) &&= \bar c(z)\left( \gamma_\alpha\gamma\cdot\stackrel{\leftrightarrow}{D} \stackrel{\leftrightarrow}{D}_{\beta} +\gamma_{\alpha}\stackrel{\leftrightarrow}{D}_{\beta}\gamma\cdot\stackrel{\leftrightarrow}{D} +\gamma_{\beta}\gamma\cdot\stackrel{\leftrightarrow}{D} \stackrel{\leftrightarrow}{D}_{\alpha} \right. \\
\notag 	&&\left.+\gamma_{\beta}\stackrel{\leftrightarrow}{D}_{\alpha}\gamma\cdot\stackrel{\leftrightarrow}{D} - g_{\alpha\beta}\gamma\cdot\stackrel{\leftrightarrow}{D} \gamma\cdot\stackrel{\leftrightarrow}{D} \right)\gamma_5c(z),\\
\notag
J^{\eta_{c2}}_{\alpha \beta }(z) &&= \bar c(z)\left({\mathord{\buildrel{\lower3pt\hbox{$\scriptscriptstyle\leftrightarrow$}} 
\over D} }_{\alpha}{\mathord{\buildrel{\lower3pt\hbox{$\scriptscriptstyle\leftrightarrow$}} 
\over D} }_{\beta}+{\mathord{\buildrel{\lower3pt\hbox{$\scriptscriptstyle\leftrightarrow$}} 
\over D} }_{\beta}{\mathord{\buildrel{\lower3pt\hbox{$\scriptscriptstyle\leftrightarrow$}} 
\over D} }_{\alpha}-\frac{1}{2}g_{\alpha\beta}\mathord{\buildrel{\lower3pt\hbox{$\scriptscriptstyle\leftrightarrow$}} 
\over{D}}\cdot\mathord{\buildrel{\lower3pt\hbox{$\scriptscriptstyle\leftrightarrow$}} 
\over{D}}\right)c(z),\\
J^{\psi_3}_{\alpha \beta \gamma}(z) &&= \bar c(z)\left( {\mathord{\buildrel{\lower3pt\hbox{$\scriptscriptstyle\leftrightarrow$}} 
\over D} }_{\alpha}{\mathord{\buildrel{\lower3pt\hbox{$\scriptscriptstyle\leftrightarrow$}} 
\over D} }_{\beta} \gamma_{\gamma}+{\mathord{\buildrel{\lower3pt\hbox{$\scriptscriptstyle\leftrightarrow$}} 
\over D} }_{\gamma}{\mathord{\buildrel{\lower3pt\hbox{$\scriptscriptstyle\leftrightarrow$}} 
\over D} }_{\alpha} \gamma_{\beta}+{\mathord{\buildrel{\lower3pt\hbox{$\scriptscriptstyle\leftrightarrow$}} 
\over D} }_{\beta}{\mathord{\buildrel{\lower3pt\hbox{$\scriptscriptstyle\leftrightarrow$}} 
\over D} }_{\gamma} \gamma_{\beta}\right)c(z),
\end{eqnarray}
where $\Gamma=\gamma_\mu$, $\gamma_\mu\gamma_5$ and $\sigma_{\mu\nu}$ or $\sigma_{\mu\nu}\gamma_5$ for vector, axial vector and tensor form factors, respectively. $\mathord{\buildrel{\lower3pt\hbox{$\scriptscriptstyle\leftrightarrow$}} 
\over{D}}_\alpha=\mathord{\buildrel{\lower3pt\hbox{$\scriptscriptstyle\rightarrow$}} \over D}_\alpha- \mathord{\buildrel{\lower3pt\hbox{$\scriptscriptstyle\leftarrow$}} 
\over D}_\alpha $, $\mathord{\buildrel{\lower3pt\hbox{$\scriptscriptstyle\rightarrow$}} 
\over D}_\alpha = \mathord{\buildrel{\lower3pt\hbox{$\scriptscriptstyle\rightarrow$}} 
\over{\partial_\alpha}}- ig_st^aG^a_\alpha$ and $\mathord{\buildrel{\lower3pt\hbox{$\scriptscriptstyle\leftarrow$}} 
\over D}_\alpha = \mathord{\buildrel{\lower3pt\hbox{$\scriptscriptstyle\leftarrow$}} 
\over{\partial_\alpha}} + ig_st^aG^a_\alpha$ represent the left and right covariant derivative, respectively.

\subsection{The phenomenological side}
To obtain the phenomenological representation of correlation function, the complete set of hadronic states which can couple to the corresponding interpolating currents is inserted into Eq.~(\ref{eq:1}). After performing the integration in coordinate space and using the double dispersion relation, the three-point correlation can be written as:
\begin{eqnarray}
\notag
\Pi^{\mathrm{phy}}(p,p')&&=\frac{\left\langle 0\right| J_X(0)\left| X(p')\right\rangle \left\langle B_c(p)\right| J^\dagger_{B_c}(0)\left|0\right\rangle}{(m_{B_c}^2-p^2)(m_X^2-p'^2)}\\
&&\times\left\langle X(p')\right|\tilde{J}(0)\left| B_c(p)\right\rangle+...,
\end{eqnarray}
where the ellipsis represents the contributions from higher resonances and continuum states. Besides, the interpolating currents of high spin meson also can couple to the low spin mesons, these redundant contributions are also represented with ellipsis. The meson vacuum matrix elements can be defined as:
\begin{eqnarray}\label{eq:4}
\notag
\left\langle 0\right|J^{\psi_1}_\alpha(0)\left|\psi_1(p')\right\rangle&&=f_{\psi_1}\epsilon^{\psi_1}_{\alpha},\\
\notag
\left\langle 0\right|J^{\psi_1}_\alpha(0)\left|\chi_{c0}(p')\right\rangle&&=f_{\chi_{c0}}p'_{\alpha},\\
\notag
\left\langle 0\right|J^{\psi_2[\eta_{c2}]}_{\alpha\beta}(0)\left|\psi_{2}(p')\right\rangle&&=f_{\psi_2[\eta_{c2}]}\epsilon^{\psi_2[\eta_{c2}]}_{\alpha\beta},\\
\notag
\left\langle 0\right|J^{\psi_2[\eta_{c2}]}_{\alpha\beta}(0)\left|\chi_{c1}(p')\right\rangle&&=f_{\chi_{c1}}(p'_{\alpha}\epsilon^{\chi_{c1}}_{\beta}+p'_{\beta}\epsilon^{\chi_{c1}}_{\alpha}),\\
\notag
\left\langle 0\right|J^{\psi_2[\eta_{c2}]}_{\alpha\beta}(0)\left|\eta_c(p')\right\rangle&&=f_{\eta_c}p'_{\alpha}p'_{\beta},\\
\notag
\left\langle 0\right|J^{\psi_3}_{\alpha\beta\gamma}(0)\left|\psi_3(p')\right\rangle&&=f_{\psi_3}\epsilon^{\psi_3}_{\alpha\beta\gamma},\\
\notag
\left\langle 0\right|J^{\psi_3}_{\alpha\beta\gamma}(0)\left|\chi_{c2}(p')\right\rangle&&=f_{\chi_{c2}}(p'_{\alpha}\epsilon^{\chi_{c2}}_{\beta\gamma}+p'_{\gamma}\epsilon^{\chi_{c2}}_{\alpha\beta}+p'_{\beta}\epsilon^{\chi_{c2}}_{\gamma\alpha}),\\
\notag
\left\langle 0\right|J^{\psi_3}_{\alpha\beta\gamma}(0)\left|J/\psi[\psi_1](p')\right\rangle&&=f_{J/\psi[\psi_1]}(p'_{\alpha}p'_{\beta}\epsilon^{J/\psi[\psi_1]}_{\gamma}\\
\notag
&&+p'_{\gamma}p'_{\alpha}\epsilon^{J/\psi[\psi_1]}_{\beta}+p'_{\beta}p'_{\gamma}\epsilon^{J/\psi[\psi_1]}_{\alpha}),\\
\notag
\left\langle 0\right|J^{\psi_3}_{\alpha\beta\gamma}(0)\left|\chi_{c0}(p')\right\rangle&&=f_{\chi_{c0}}p'_{\alpha}p'_{\beta}p'_{\gamma},\\
\left\langle B_c(p)\right|J_{B_c}^\dagger(0)\left|0\right\rangle&&=\frac{f_{B_c}m_{B_c}^2}{m_b+m_c},
\end{eqnarray}
where $f_{\psi_1}$, $f_{\chi_{c0}}$, $f_{\psi_2}$, $f_{\eta_{c2}}$, $f_{\chi_{c1}}$, $f_{\eta_c}$, $f_{\psi_3}$, $f_{\chi_{c2}}$, $f_{J/\psi}$ and $f_{B_c}$ are the decay constants of corresponding mesons, $\epsilon_{\alpha}$, $\epsilon_{\alpha\beta}$ and $\epsilon_{\alpha\beta\gamma}$ denote the polarization tensors. It can be seen from Eq.~(\ref{eq:4}) that the current of $\psi_1$ ($J^P=1^-$) can couple to $\chi_{c0}$ ($J^P=0^+$), the currents of $\psi_2$ and $\eta_{c2}$ ($J^P=2^-$) can couple to both $\chi_{c1}$ ($J^P=1^+$) and $\eta_{c}$ ($J^P=0^-$), and the current of $\psi_3$ ($J^P=3^-$) can couple to $\chi_{c2}$ ($J^P=2^+$), $J/\psi[\psi_1]$ ($J^P=1^-$) and $\chi_{c0}$ ($J^P=0^+$). To eliminate the contamination of the redundant coupling, the projection operator should be introduced in corresponding correlation functions. The hadron transition matrix elements can be defined as in terms of form factors~\cite{Chen:2017vgi,Wirbel:1985ji}:
\begin{widetext}
\begin{eqnarray}\label{eq:5}
\notag
\left\langle \psi _1(p') \right|\bar c(0)\gamma _\mu b(0)\left| B_c(p) \right\rangle  &&=  g^{B_c \to \psi _1}(q^2)\varepsilon _{\mu \lambda \tau \sigma }\epsilon ^{\psi _1*\lambda }P^\tau q^\sigma,\\
\notag
\left\langle \psi _1(p') \right|\bar c(0)\gamma _\mu \gamma _5b(0)\left| B_c(p) \right\rangle  &&= - i\left\{f^{B_c \to \psi _1}(q^2)\epsilon _{\mu}^{\psi _1*}+\epsilon^{\psi _2*}\cdot P \left[a_ + ^{B_c \to \psi _1}(q^2)P_\mu + a_ - ^{B_c \to \psi _1}(q^2)q_\mu\right]\right\},\\
\notag
\left\langle \psi _1(p') \right|\bar c(0)\sigma _{\mu \nu }b(0)\left| B_c(p) \right\rangle &&=i\left[ T_0^{B_c \to \psi _1}(q^2)\frac{\epsilon ^{\psi _1*} \cdot q}{(m_{B_c}+ m_{\psi _1})^2} \varepsilon _{\mu \nu \lambda \tau}p^\lambda p'^\tau+ T_1^{B_c \to \psi _1}(q^2)\varepsilon _{\mu \nu \lambda \tau}p^\lambda\epsilon^{\psi _1*\tau}+ T_2^{B_c \to \psi _1}(q^2)\varepsilon _{\mu \nu \lambda \tau }p'^\lambda \epsilon ^{\psi _1*\tau }\right],\\
\end{eqnarray}
\begin{eqnarray}
\notag
\left\langle \psi _2[\eta _{c2}](p') \right|\bar c(0)\gamma _\mu b(0)\left| B_c(p) \right\rangle  &&= i\left\{ m^{B_c \to \psi _2[\eta _{c2}]}(q^2)\epsilon_{\mu \lambda}^{\psi _2[\eta _{c2}]*}P^\lambda+\epsilon^{\psi _2[\eta _{c2}]*}_{\lambda \tau}P^\lambda P^\tau\left[z_ + ^{B_c \to \psi _2[\eta _{c2}]}(q^2)P_\mu + z_ - ^{B_c \to \psi _2[\eta _{c2}]}(q^2)q_\mu\right]\right\},\\
\notag
\left\langle \psi _2[\eta _{c2}](p') \right|\bar c(0)\gamma _\mu \gamma _5b(0)\left| B_c(p) \right\rangle  &&= -n^{B_c \to \psi _2[\eta _{c2}]}(q^2)\varepsilon _{\mu \tau \rho \sigma}\epsilon_{\tau \lambda }^{\psi _2[\eta _{c2}]*}P^\lambda P^\rho q^\sigma ,\\
\notag
\left\langle \psi _2[\eta _{c2}](p') \right|\bar c(0)\sigma _{\mu \nu }\gamma _5b(0)\left| B_c(p) \right\rangle  &&= -\frac{iq^\sigma}{m_{B_c}}\left[ {T_0^{B_c \to \psi _2[\eta _{c2}]}(q^2)\frac{\epsilon _{\rho \sigma }^{\psi _2[\eta _{c2}]*}q^\rho }{(m_{B_c} + m_{\psi _2[\eta _{c2}]})^2}\varepsilon _{\mu \nu \lambda \tau }p^\lambda p'^\tau + T_1^{B_c \to \psi _2[\eta _{c2}]}(q^2)\varepsilon_{\mu \nu \lambda \tau }\epsilon_{\lambda \sigma }^{\psi _2[\eta _{c2}]*}p_\tau } \right.\\
&&\left. { + T_2^{B_c \to \psi _2[\eta _{c2}]}(q^2)\varepsilon _{\mu \nu \lambda \tau }\epsilon _{\lambda \sigma }^{\psi _2[\eta _{c2}]*}p'_\tau} \right],
\end{eqnarray}
\begin{eqnarray}\label{eq:7}
\notag
\left\langle \psi _3(p') \right|\bar c(0)\gamma _\mu b(0)\left| B_c(p) \right\rangle  &&= y^{B_c \to \psi _3}(q^2)\varepsilon _{\mu \lambda \tau \rho }\epsilon _{\lambda \chi \sigma }^{\psi _3*}P_\chi P_\sigma P^\tau q^\rho,\\
\notag
\left\langle \psi _3(p') \right|\bar c(0)\gamma _\mu \gamma _5b(0)\left| B_c(p) \right\rangle  &&=  - i\left\{w^{B_c \to \psi _3}(q^2)\epsilon _{\mu \lambda \tau }^{\psi _3*}P^\lambda P^\tau +\epsilon _{\lambda \tau \rho }^{\psi _3*}P^\lambda P^\tau P^\rho \left[o_ + ^{B_c \to \psi _3}(q^2)P_\mu + o_ - ^{B_c \to \psi _3}(q^2)q_\mu\right]\right\},\\
\notag
\left\langle \psi _3(p') \right|\bar c(0)\sigma _{\mu \nu }b(0)\left| B_c(p) \right\rangle  &&= \frac{iq^\sigma q^\chi}{m_{B_c}^2}\left[ {T_0^{B_c \to \psi _3}(q^2)\frac{\varepsilon _{\rho \sigma \chi }^{\psi _3*}q^\rho}{(m_{B_c} + m_{\psi _3})^2}\varepsilon _{\mu \nu \lambda \tau }p^\lambda p'^\tau + T_1^{B_c \to \psi _3}(q^2)\varepsilon _{\mu \nu \lambda \tau }\epsilon _{\lambda \sigma \chi }^{\psi _3*}p_\tau } \right.\\
&&\left. { + T_2^{B_c \to \psi _3}(q^2)\varepsilon _{\mu \nu \lambda \tau }\epsilon _{\lambda \sigma \chi }^{\psi _3*}p'_\tau } \right],
\end{eqnarray}
\end{widetext}
where $P=p+p'$ and $q=p-p'$. $g$ ($f/a_+/a_-$), $m/z_+/z_-$ ($n$) and $y$ ($w/o_+/o_-$) are the vector (axial vector) form factors of $\psi_1$, $\psi_2[\eta_{c2}]$ and $\psi_3$, $T_0/T_1/T_2$ are the tensor form factors. 

According these above equations, the phenomenological representation of correlation functions can be obtained and decomposed into different tensor structures. The form factors are included in these tensor structures.
\subsection{The QCD side}
\begin{figure}
	\centering
	\includegraphics[width=8.5cm]{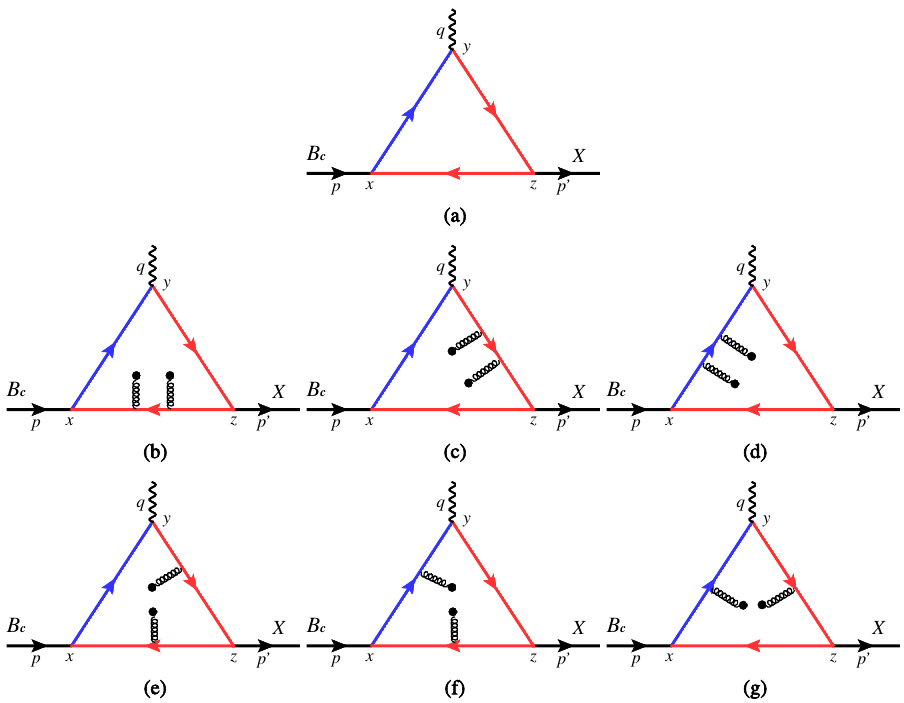}
	\caption{The Feynman diagrams for the perturbative part (a) and gluon condensate (b)-(g), where the blue and red solid lines denote the $b$ and $c$ quark lines, respectively. The black loop lines are the gluon
		lines}
	\label{FDQ}
\end{figure}

After contracting the quark filed with Wick’s theorem and doing the operator product expansion (OPE) of correlation function, the QCD representation of the correlation function can be written as,
\begin{widetext}
\begin{eqnarray}\label{eq:8}
\notag
\Pi _{1\alpha \mu }^{\psi _1\mathrm{QCD}}(p,p') &&=  i\int {d^4xd^4ye^{ip'\cdot z}e^{i(p - p')\cdot y}e^{ - ip\cdot x}Tr[C^{km}(x - z)\Gamma _\alpha (z)C^{mn}(z - y)\gamma _\mu B^{nk}(y - x)\gamma _5]} \big{|}_{z \to 0},\\
\notag
\Pi _{2\alpha \mu }^{\psi _1\mathrm{QCD}}(p,p') &&= i\int {d^4xd^4ye^{ip'\cdot z}e^{i(p - p')\cdot y}e^{ - ip\cdot x}Tr[C^{km}(x - z)\Gamma _\alpha (z)C^{mn}(z - y)\gamma _\mu \gamma _5B^{nk}(y - x)\gamma _5]} \big{|}_{z \to 0},\\
\Pi _{\alpha \mu \nu }^{\psi _1\mathrm{QCD}}(p,p') &&= i\int {d^4xd^4ye^{ip'\cdot z}e^{i(p - p')\cdot y}e^{ - ip\cdot x}Tr[C^{km}(x - z)\Gamma _\alpha (z)C^{mn}(z - y)\sigma _{\mu \nu }B^{nk}(y - x)\gamma _5]} \big{|}_{z \to 0},
\end{eqnarray}
\begin{eqnarray}
\notag
\Pi _{1\alpha \beta \mu }^{\psi _2[\eta_{c2}]\mathrm{QCD}}(p,p') &&=  i\int {d^4xd^4ye^{ip'\cdot z}e^{i(p - p')\cdot y}e^{ - ip\cdot x}} Tr[C^{km}(x - z)\Gamma _{\alpha \beta }^{1(2)}(z)C^{mn}(z - y)\gamma _\mu B^{nk}(y - x)\gamma _5]\big{|}_{z \to 0},\\
\notag
\Pi _{2\alpha \beta \mu }^{\psi _2[\eta_{c2}]\mathrm{QCD}}(p,p') &&=  i\int {d^4xd^4ye^{ip'\cdot z}e^{i(p - p')\cdot y}e^{ - ip\cdot x}} Tr[C^{km}(x - z)\Gamma _{\alpha \beta }^{1(2)}(z)C^{mn}(z - y)\gamma _\mu \gamma _5B^{nk}(y - x)\gamma _5]\big{|}_{z \to 0},\\
\Pi _{\alpha \beta \mu \nu }^{\psi _2[\eta_{c2}]\mathrm{QCD}}(p,p') &&=  i\int {d^4xd^4ye^{ip'\cdot z}e^{i(p - p')\cdot y}e^{ - ip\cdot x}} Tr[C^{km}(x - z)\Gamma _{\alpha \beta }^{1(2)}(z)C^{mn}(z - y)\sigma _{\mu \nu }\gamma_5B^{nk}(y - x)\gamma _5]\big{|}_{z \to 0},
\end{eqnarray}
\begin{eqnarray}\label{eq:10}
\notag
\Pi _{1\alpha \beta \gamma \mu }^{\psi _3\mathrm{QCD}}(p,p') &&=  i\int {d^4xd^4ye^{ip'\cdot z}e^{i(p - p')\cdot y}e^{ - ip\cdot x}} Tr[C^{km}(x - z)\Gamma _{\alpha \beta \gamma}(z)C^{mn}(z - y)\gamma _\mu B^{nk}(y - x)\gamma _5]\big{|}_{z \to 0},\\
\notag
\Pi _{2\alpha \beta \gamma \mu }^{\psi _3\mathrm{QCD}}(p,p') &&=  i\int {d^4xd^4ye^{ip'\cdot z}e^{i(p - p')\cdot y}e^{ - ip\cdot x}} Tr[C^{km}(x - z)\Gamma _{\alpha \beta \gamma }(z)C^{mn}(z - y)\gamma _\mu \gamma _5B^{nk}(y - x)\gamma _5]\big{|}_{z \to 0}\\
\Pi _{\alpha \beta \gamma \mu \nu }^{\psi _3\mathrm{QCD}}(p,p') &&=  i\int {d^4xd^4ye^{ip'\cdot z}e^{i(p - p')\cdot y}e^{ - ip\cdot x}} Tr[C^{km}(x - z)\Gamma _{\alpha \beta \gamma }(z)C^{mn}(z - y)\sigma _{\mu \nu }B^{nk}(y - x)\gamma _5]\big{|}_{z \to 0},
\end{eqnarray}
\end{widetext}
where the vertex factors $\Gamma$ can be expressed as,
\begin{eqnarray}
\notag
\Gamma _\alpha (z) &&= {{\mathord{\buildrel{\lower3pt\hbox{$\scriptscriptstyle\leftrightarrow$}} 
	\over D} }_{\alpha'} }{{\mathord{\buildrel{\lower3pt\hbox{$\scriptscriptstyle\leftrightarrow$}} 
\over D} }_{\beta'} }{\gamma _\rho }({g^{\alpha' \beta' }}{g^{\rho \alpha }} + {g^{\alpha' \rho }}{g^{\beta' \alpha }} + {g^{\rho \beta' }}{g^{\alpha' \alpha }}),\\
\notag
\Gamma _{\alpha \beta }^1(z) &&= {\gamma _\alpha }\gamma  \cdot \mathord{\buildrel{\lower3pt\hbox{$\scriptscriptstyle\leftrightarrow$}} 
		\over D} {{\mathord{\buildrel{\lower3pt\hbox{$\scriptscriptstyle\leftrightarrow$}} 
				\over D} }_\beta } + {\gamma _\alpha }{{\mathord{\buildrel{\lower3pt\hbox{$\scriptscriptstyle\leftrightarrow$}} 
				\over D} }_\beta }\gamma  \cdot \mathord{\buildrel{\lower3pt\hbox{$\scriptscriptstyle\leftrightarrow$}} 
		\over D} + {\gamma _\beta }\gamma  \cdot \mathord{\buildrel{\lower3pt\hbox{$\scriptscriptstyle\leftrightarrow$}} 
		\over D} {{\mathord{\buildrel{\lower3pt\hbox{$\scriptscriptstyle\leftrightarrow$}} 
				\over D} }_\alpha } \\
\notag
&& + {\gamma _\beta }{{\mathord{\buildrel{\lower3pt\hbox{$\scriptscriptstyle\leftrightarrow$}} 
				\over D} }_\alpha }\gamma  \cdot \mathord{\buildrel{\lower3pt\hbox{$\scriptscriptstyle\leftrightarrow$}} 
		\over D}  - {g_{\alpha \beta }}\gamma  \cdot \mathord{\buildrel{\lower3pt\hbox{$\scriptscriptstyle\leftrightarrow$}} 
		\over D} \gamma  \cdot \mathord{\buildrel{\lower3pt\hbox{$\scriptscriptstyle\leftrightarrow$}} 
		\over D}, \\
\notag
\Gamma _{\alpha \beta }^2(z) &&= {{\mathord{\buildrel{\lower3pt\hbox{$\scriptscriptstyle\leftrightarrow$}} 
				\over D} }_\alpha }{{\mathord{\buildrel{\lower3pt\hbox{$\scriptscriptstyle\leftrightarrow$}} 
				\over D} }_\beta } + {{\mathord{\buildrel{\lower3pt\hbox{$\scriptscriptstyle\leftrightarrow$}} 
				\over D} }_\beta }{{\mathord{\buildrel{\lower3pt\hbox{$\scriptscriptstyle\leftrightarrow$}} 
				\over D} }_\alpha } - \frac{1}{2}{g_{\alpha \beta }}\mathord{\buildrel{\lower3pt\hbox{$\scriptscriptstyle\leftrightarrow$}} 
		\over D}\cdot \mathord{\buildrel{\lower3pt\hbox{$\scriptscriptstyle\leftrightarrow$}} 
		\over D}, \\
{\Gamma _{\alpha \beta \gamma }}(z) &&= {{\mathord{\buildrel{\lower3pt\hbox{$\scriptscriptstyle\leftrightarrow$}} 
				\over D} }_\alpha }{{\mathord{\buildrel{\lower3pt\hbox{$\scriptscriptstyle\leftrightarrow$}} 
				\over D} }_\beta }{\gamma _\gamma } + {{\mathord{\buildrel{\lower3pt\hbox{$\scriptscriptstyle\leftrightarrow$}} 
				\over D} }_\gamma }{{\mathord{\buildrel{\lower3pt\hbox{$\scriptscriptstyle\leftrightarrow$}} 
				\over D} }_\alpha }{\gamma _\beta } + {{\mathord{\buildrel{\lower3pt\hbox{$\scriptscriptstyle\leftrightarrow$}} 
				\over D} }_\beta }{{\mathord{\buildrel{\lower3pt\hbox{$\scriptscriptstyle\leftrightarrow$}} 
				\over D} }_\gamma }{\gamma _\alpha }.
\end{eqnarray}
$C^{ij}(x)$ and $B^{ij}(x)$ are the full propagator of $c$ and $b$ quarks, and can be uniformly represented as the following forms~\cite{Reinders:1984sr}:
\begin{eqnarray}
	\notag
	Q^{ij}(x) &&= \frac{i}{(2\pi )^4}\int d^4 k e^{ - ik \cdot x} \left \{\frac{\delta ^{ij}}{\slashed k - {m_Q}} \right. \\
	\notag
	&& - \frac{g_sG_{\alpha \beta }^nt_{ij}^n}{4}\frac{\sigma ^{\alpha \beta }(\slashed{k} + {m_Q}) + (\slashed{k} + {m_Q})\sigma ^{\alpha \beta }}{(k^2 - m_Q^2)^2}\\
	\notag
	&& \left. - \frac{g_s^2({t^a}{t^b})_{ij}G_{\alpha \beta }^aG_{\mu \nu }^b(f^{\alpha \beta \mu \nu } + f^{\alpha \mu \beta \nu } + f^{\alpha \mu \nu \beta })}{4(k^2 - m_Q^2)^5} + ... \right\}. \\
\end{eqnarray}
Here $Q$ denotes $C$ or $B$, $t^{n}=\frac{\lambda^{n}}{2}$, $\lambda^{n}(n=1,...,8)$ are the Gell-Mann matrices, $i$ and $j$ are color indices, $\sigma_{\alpha\beta}=\frac{i}{2}[\gamma_{\alpha},\gamma_{\beta}]$, and $f^{\alpha\beta\mu\nu}$ have the following form:
\begin{eqnarray}
	\notag
	f^{\alpha \beta \mu \nu }&&= (\slashed k + m_Q)\gamma ^\alpha (\slashed k + m_Q)\gamma ^\beta (\slashed k + m_Q)\\
	&& \times \gamma ^\mu (\slashed k + m_Q)\gamma ^\nu (\slashed k + m_Q).
\end{eqnarray}
The gluon filed $G^a_\mu$ in the covariant derivative has no contributions for correlation functions since we take the fixed point (Schwinger) gauge~\cite{Yu:2021ggd}. In this gauge, the gluon filed can be written as:
\begin{eqnarray}
G^a_\mu(z)=\frac{1}{2}z^\nu G^a_{\nu\mu}+\frac{1}{3}z^\nu z^\rho \frac{\partial G^a_{\nu\mu}}{\partial z^\rho}+...,
\end{eqnarray}
this term will disappear when $z\to 0$. Substituting with the full propagator in Eqs.~(\ref{eq:8})-(\ref{eq:10}), the correlation functions in QCD side can be obtained and also decomposed into following different tensor structures:
\begin{eqnarray}\label{eq:15}
\notag
\tilde \Pi _{1\xi \mu }^{\psi _1\mathrm{QCD}}(p,p') &&= \left( g^{\xi \alpha } - \frac{p'^\xi p'^\alpha }{p'^2} \right)\Pi _{1\alpha \mu }^{\psi _1\mathrm{QCD}}\\
\notag
&&= \Pi _1^{\psi _1\mathrm{QCD}}\varepsilon _{\xi \mu \lambda \tau }p^\lambda p'^\tau,\\
\notag
\tilde \Pi _{2\xi \mu }^{\psi _1\mathrm{QCD}}(p,p') &&= \left( g^{\xi \alpha } - \frac{p'^\xi p'^\alpha }{p'^2} \right)\Pi _{2\alpha \mu }^{\psi _1\mathrm{QCD}}\\
\notag
&&= \Pi _2^{\psi _1\mathrm{QCD}}g_{\xi \mu } + \Pi _3^{\psi _1\mathrm{QCD}}p^\xi p^\mu \\
\notag
&&+ \Pi _4^{\psi _1\mathrm{QCD}}p^\xi p'^\mu + ...,\\
\notag
\tilde \Pi _{\xi \mu \nu }^{\psi _1\mathrm{QCD}}(p,p') &&= \left( g^{\xi \alpha }- \frac{p'^\xi p'^\alpha }{p'^2} \right)\Pi _{\alpha \mu \nu }^{\psi _1\mathrm{QCD}}\\
\notag
&&= \Pi _5^{\psi _1\mathrm{QCD}}p_\xi\varepsilon _{\mu \nu \lambda \tau }p^\lambda p'^\tau + \Pi _6^{\psi _1\mathrm{QCD}}\varepsilon _{\xi \mu \nu \lambda }p^\lambda\\
&&+ \Pi _7^{\psi _1\mathrm{QCD}}\varepsilon _{\xi \mu \nu \lambda }p'^\lambda + ...,
\end{eqnarray}
\begin{eqnarray}\label{eq:16}
\notag
\tilde \Pi _{1\xi \psi \mu }^{\psi _2[\eta_{c2}]\mathrm{QCD}}(p,p') &&= \left( g^{\xi \alpha } - \frac{p'^\xi p'^\alpha }{p'^2} \right)\left( g^{\psi \beta } - \frac{p'^\psi p'^\beta}{p'^2} \right)\Pi _{1\alpha \beta \mu}^{\psi _2[\eta_{c2}]\mathrm{QCD}}\\
\notag
&&= \Pi _1^{\psi _2[\eta_{c2}]\mathrm{QCD}}g_{\mu \psi }p_\xi + \Pi _2^{\psi _2[\eta_{c2}]\mathrm{QCD}}p_\mu p_\psi p_\xi\\
\notag
&&+ \Pi _3^{\psi _2[\eta_{c2}]\mathrm{QCD}}p'_\mu p_\psi p_\xi + ...,\\
\notag
\tilde \Pi _{2\xi \psi \mu }^{\psi _2[\eta_{c2}]\mathrm{QCD}}(p,p') &&= \left( g^{\xi \alpha } - \frac{p'^\xi p'^\alpha }{p'^2} \right)\left( g^{\psi \beta } - \frac{p'^\psi p'^\beta}{p'^2} \right)\Pi _{2\alpha \beta \mu }^{\psi _2[\eta_{c2}]\mathrm{QCD}}\\
\notag
&&= \Pi _4^{\psi _2[\eta_{c2}]\mathrm{QCD}}p_\xi\varepsilon _{\mu \psi \lambda \tau }p^\lambda p'^\tau + ...,\\
\notag
\tilde \Pi _{\xi \psi \mu \nu }^{\psi _2[\eta_{c2}]\mathrm{QCD}}(p,p') &&= \left( g^{\xi \alpha } - \frac{p'^\xi p'^\alpha}{p'^2} \right)\left( g^{\psi \beta } - \frac{p'^\psi p'^\beta}{p'^2} \right)\Pi _{\alpha \beta \mu \nu }^{\psi _2[\eta_{c2}]\mathrm{QCD}}\\
\notag
&&= \Pi _5^{\psi _2[\eta_{c2}]\mathrm{QCD}}\varepsilon _{\mu \nu \lambda \tau }p^\lambda p'^\tau p_\psi p_\xi \\
\notag
&& + \Pi _6^{\psi _2[\eta_{c2}]\mathrm{QCD}}\varepsilon _{\mu \nu \psi \lambda }p^\lambda p_\xi \\
&&+ \Pi _7^{\psi _2[\eta_{c2}]\mathrm{QCD}}\varepsilon _{\mu \nu \psi \lambda }p'^\lambda p_\xi  + ...,
\end{eqnarray}
\begin{eqnarray}\label{eq:17}
\notag
\tilde \Pi _{1\xi \psi \epsilon \mu }^{\psi _3\mathrm{QCD}}(p,p') &&= \left( g^{\xi \alpha }- \frac{p'^\xi p'^\alpha}{p'^2} \right)\left( g^{\psi \beta } - \frac{p'^\psi p'^\beta }{p'^2} \right)\\
\notag
&&\times \left( g^{\epsilon \gamma } - \frac{p'^\epsilon p'^\gamma}{p'^2} \right)\Pi _{1\alpha \beta \gamma \mu }^{\psi _3\mathrm{QCD}}\\
\notag
&&= \Pi _1^{\psi _3\mathrm{QCD}}\varepsilon _{\mu \psi \lambda \tau }p^\lambda p'^\tau p_\xi p_\epsilon + ...,\\
\notag
\tilde \Pi _{2\xi \psi \epsilon \mu }^{\psi _3\mathrm{QCD}}(p,p') &&= \left( g^{\xi \alpha } - \frac{p'^\xi p'^\alpha}{p'^2} \right)\left( g^{\psi \beta } - \frac{p'^\psi p'^\beta }{p'^2} \right)\\
\notag
&&\times \left( g^{\epsilon \gamma } - \frac{p'^\epsilon p'^\gamma}{p'^2} \right)\Pi _{2\alpha \beta \gamma \mu }^{\psi _3\mathrm{QCD}}\\
\notag
&&= \Pi _2^{\psi _3\mathrm{QCD}}g_{\mu \xi }p_\epsilon p_\psi + \Pi _3^{\psi _3\mathrm{QCD}}p_\mu p_\xi p_\epsilon p_\psi\\
\notag
&&+ \Pi _4^{\psi _3\mathrm{QCD}}p'_\mu p_\xi p_\epsilon p_\psi + ...,\\
\notag
\tilde \Pi _{\xi \psi \epsilon \mu \nu }^{\psi _3\mathrm{QCD}}(p,p') &&= \left( g^{\xi \alpha } - \frac{p'^\xi p'^\alpha}{p'^2} \right)\left( g^{\psi \beta } - \frac{p'^\psi p'^\beta}{p'^2} \right)\\
\notag
&&\times \left( g^{\epsilon \gamma } - \frac{p'^\epsilon p'^\gamma}{p'^2} \right)\Pi _{\alpha \beta \gamma \mu \nu }^{\psi _3\mathrm{QCD}}\\
\notag
&&= \Pi _5^{\psi _3\mathrm{QCD}}p_\psi p_\xi p_\epsilon \varepsilon _{\mu \nu \lambda \tau }p^\lambda p'^\tau\\
\notag
&&+ \Pi _6^{\psi _3\mathrm{QCD}}p_\xi p_\epsilon \varepsilon _{\mu \nu \psi \lambda }p^\lambda\\
&& + \Pi _7^{\psi _3\mathrm{QCD}}p_\xi p_\epsilon \varepsilon _{\mu \nu \psi \lambda }p'^\lambda + ....
\end{eqnarray}
The projection operator $\left( g^{\xi \alpha} - \frac{p'^\xi p'^\alpha }{p'^2} \right)$ in Eq.~(\ref{eq:15}) is introduced to eliminate the coupling of the current $J_{\alpha}^{\psi_1}$ with $\chi_{c0}$, $\left( g^{\xi \alpha} - \frac{p'^\xi p'^\alpha }{p'^2} \right)\left( g^{\psi \beta} - \frac{p'^\psi p'^\beta }{p'^2} \right)$ in Eq.~(\ref{eq:16}) is to eliminate the coupling of the current $J_{\alpha\beta}^{\psi_2[\eta_{c2}]}$ with $\chi_{c1}$ and $\eta_c$ and $\left( g^{\xi \alpha} - \frac{p'^\xi p'^\alpha }{p'^2} \right)\left( g^{\psi \beta} - \frac{p'^\psi p'^\beta }{p'^2} \right)\left( g^{\epsilon \gamma} - \frac{p'^\epsilon p'^\gamma }{p'^2} \right)$ in Eq.~(\ref{eq:17}) is to eliminate the coupling of the current $J_{\alpha\beta\gamma}^{\psi_3}$ with $\chi_{c2}$, $J/\psi[\psi_1]$ and $\chi_{c0}$. $\Pi^{\mathrm{QCD}}_i$ in the right side of these above equations are the scalar invariant amplitudes which include perturbative and non-perturbative parts,
\begin{eqnarray}
\Pi^{\mathrm{QCD}}_i(p,p')=\Pi^{\mathrm{pert}}_i(p,p')+\Pi^{\mathrm{non-pert}}_i(p,p').
\end{eqnarray}
Since the heavy quark will not contribute to quark condensate, the main non-perturbative contributions are two gluon condensate $\langle g_s^2GG\rangle$, three gluon condensate $g_s^3\langle fGGG\rangle$ and higher dimension gluon condensate terms. According our previous work, the contributions from three gluon condensate is about one-tenth that of the two gluon condensate~\cite{Wu:2024gcq}. In addition, the higher dimension gluon condensate terms are further suppressed by $O(\alpha_s)$ ($\alpha_s=g_s^2/4\pi$). Thus, we only reserve the two gluon condensate term in our calculation. The corresponding Feynman diagrams are shown in Fig.~\ref{FDQ}. By using the double dispersion relation, the scalar invariant amplitude can be expressed as the following form~\cite{Colangelo:2000dp}:
\begin{eqnarray}
\notag
\Pi^{\mathrm{QCD}}_i(p,p')&&=\int\limits_{s_{\min}}^{\infty}\int\limits_{u_{\min}}^{\infty}dsdu\frac{\rho^{\mathrm{QCD}}_i(s,u,q^2)}{(s-p^2)(u-p'^2)},\\
\rho^{\mathrm{QCD}}_i(s,u,q^2)&&=\rho^{\mathrm{pert}}_i(s,u,q^2)+\rho^{\langle g_s^2GG\rangle}_{i}(s,u,q^2),
\end{eqnarray}
where $\rho^{\mathrm{QCD}}(s,u,q^2)$ denote the QCD spectral density with $s=p^2$, $u=p'^2$ and $q=p-p'$. $s_{\min}$ and $u_{\min}$ are the creative thresholds for the $B_c$ meson and $D$-wave charmonium, and take the values of $(m_b+m_c)^2$ and $4m_c^2$, respectively.

The QCD spectral density of the perturbative part and two gluon condensate can both obtained by Cutkoskys's rules~\cite{Cutkosky:1960sp}, the calculation details can be found in Ref.~\cite{Wu:2024gcq}. After taking the variables change $p^2\to-P^2$, $p'^2\to-P'^2$ and $q^2\to-Q^2$, we can perform the double Borel transforms~\cite{Reinders:1984sr} for variables $P^2$ and $P'^2$ to both phenomenological and QCD sides. The variables $P^2$ and $P'^2$ will be replaced by $T_1^2$ and $T_2^2$ which are called Borel parameters. For simplicity, the relations $T^2=T_1^2$ and $T_2^2=kT_1^2$ are adopted to reduce the two Borel parameters, where $k=m_X^2/m_{B_c}^2$ and $X$ represents the $D$-wave charmonium. The plausibility for this reduction was discussed in Ref.~\cite{Lu:2025bvi}. By matching the calculations for phenomenological and QCD sides and using quark-hadron duality condition, the QCDSR for the form factors can be obtained. For simplicity, we take the vector form factor of $B_c\to\psi_1$ as an example. This form factor can be obtained by analyzing the invariant amplitude $\Pi _1^{\psi _1\mathrm{QCD}}$ in Eq.~(\ref{eq:15}) and can be expressed as:
\begin{eqnarray}\label{eq:20}
\notag
g^{B_c \to \psi_1}(Q^2) &&=\frac{m_b + m_c}{2f_{\psi _1}f_{B_c}m_{B_c}^2}\exp \left( \frac{m_{B_c}^2}{T^2} + \frac{m_{\psi _1}^2}{kT^2} \right)\\
\notag
&&\times \int\limits_{{s_{\min }}}^{s_0} \int\limits_{{u_{\min }}}^{u_0} dsdu \rho _1^{\psi _1\mathrm{QCD}}(s,u,Q^2)\exp \left(- \frac{s}{T^2} - \frac{u}{kT^2} \right),\\
\end{eqnarray}
where $s_0$ and $u_0$ are the threshold parameters for $B_c$ meson and $D$-wave charmonium which are induced to eliminate the contributions from higher resonances and continuum states. They often fulfill the relations $s_0=(m_{B_c}+\delta)^2$ and $u_0=(m_{X}+\delta)^2$, where $\delta$ denotes the energy gap between the ground and first excited states and commonly takes the value of $0.4-0.6$ GeV~\cite{Bracco:2011pg}. The full expressions of the sum rules for other form factors are shown in Appendix \ref{Sec:AppA}.
\section{Semileptonic decay of $B_c$ to $D$-wave charmonia}\label{sec3}
\begin{figure}
	\centering
	\includegraphics[width=8.5cm]{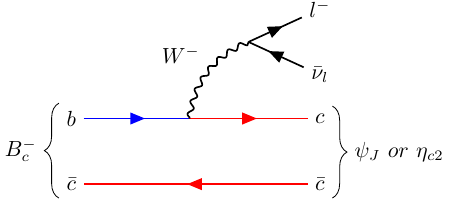}
	\caption{Feynman diagram of semileptonic decay $B_c\to\psi_{J}[\eta_{c2}]l\bar{\nu}_l$.}
	\label{FDH}
\end{figure}

The semileptonic decay $B_c \to Xl\bar{\nu}_l$ ($l=e,\mu$ and $\tau$) is dominated by the transition $b \to cl\bar{\nu}_l$ at the quark level. The corresponding Feynman diagram is shown as Fig. \ref{FDH}. This process can be described by the following effective Hamiltonian:
\begin{eqnarray}
	H_{eff} = \frac{G_F}{\sqrt 2 }V_{cb}\bar c\gamma _\mu (1 - \gamma _5)b\bar{ v}_l\gamma _\mu (1 - \gamma _5)l,
\end{eqnarray}
where $G_F=1.16637\times10^{-5}$ GeV$^{-2}$ is the Fermi constant and $V_{cb}$ is the CKM matrix element. 

With above Hamiltonian, the transition matrix element of
these decay processes can be written as:
\begin{eqnarray}
	\notag
	T &&= \left\langle X(p')l(l)\bar{\nu}_l(v)\right|H_{eff}\left| B_c(p) \right\rangle \\
	\notag
	&&=\frac{G_F}{\sqrt 2}V_{cb}\left\langle X(p') \right|\bar c\gamma _\mu (1 - \gamma _5)b\left| B_c(p) \right\rangle\\
	&&\times \left\langle l(l)\bar{\nu}_l(v)\right|\bar v_l\gamma _\mu(1 - \gamma _5)l\left| 0 \right\rangle,
\end{eqnarray}
where the leptonic part can be formulated as the following form by electroweak perturbation theory
\begin{eqnarray}
	\langle l(l)\bar{\nu}_l(v) |\bar v_l\gamma _\mu(1 - \gamma _5)l | 0 \rangle  = \bar u_{v,s}\gamma _\mu(1 - \gamma _5)u_{ - l,s'}.
\end{eqnarray}
Here $u_{v,s}$ and $u_{-l,s'}$ are the spinor wave functions of $\bar{\nu}_l$ and $l$ with momentum (spin) $l$ and $v$ ($s'$ and $s$), respectively. The hadronic part can be parameterized as the vector and axial vector form factors in Eqs.~(\ref{eq:5})-(\ref{eq:7}).

\section{Numerical results and discussions}\label{sec4}
The input parameters in this analysis are all listed in Table~\ref{IP}. The masses of heavy quarks are energy dependent, which can be expressed as the following renormalization group equation:
\begin{eqnarray}
\notag
m_Q(\mu ) &&= m_Q(m_Q)\left[ \frac{\alpha _s(\mu )}{\alpha _s(m_Q)} \right]^{\frac{12}{33 - 2n_f}},\\
\notag
\alpha _s(\mu ) &&= \frac{1}{b_0 t}\left[1 - \frac{b_1}{b_0^2}\frac{\log t}{t} \right.\\
&&\left. + \frac{b_1^2({\log^2}t - \log t - 1) + b_0b_2}{b_0^4t^2} \right],
\end{eqnarray}
where $t=\log\left(\frac{\mu^{2}}{\Lambda_{\mathrm{QCD}}^{2}}\right)$, $b_{0}=\frac{33-2n_{f}}{12\pi}$, $b_{1}=\frac{153-19n_{f}}{24\pi^{2}}$ and $b_{2}=\frac{2857-\frac{5033}{9}n_{f}+\frac{325}{27}n_{f}^{2}}{128\pi^{3}}$. $\Lambda_{\mathrm{QCD}}=213$ MeV for the quark flavors $n_{f}=5$ in the present work~\cite{ParticleDataGroup:2024cfk}. The minimum subtraction masses of $c$ and $b$ quarks are taken from the Particle Date Group~\cite{ParticleDataGroup:2024cfk}, which are $m_{c}(m_{c})=1.275\pm0.025$ GeV and $m_{b}(m_{b})=4.18\pm0.03$ GeV. Based on our previous work~\cite{Wang:2024fwc}, the energy scale $\mu=2$ GeV works well in $B_{c}$ meson system. Thus, this value is still adopted in this analysis.
\begin{table}[htbp]
	\renewcommand\arraystretch{1.3}
	\begin{ruledtabular}\caption{Input parameters (IP) used in present work.}
		\label{IP}
		\begin{tabular}{c c c c }
			IP&Values (GeV)&IP&Values \\ \hline
			$m_{B_c}$&$6.274$~\cite{ParticleDataGroup:2024cfk}&$f_{B_c}$&$0.371\pm0.037$ GeV~\cite{Wang:2024fwc}\\
			$m_{\psi_1}$&$3.77$~\cite{ParticleDataGroup:2024cfk}&$f_{\psi_{1}}$&$13.82^{+2.18}_{-1.98}$ GeV$^4$~\cite{Xin:2024xbp} \\
			$m_{\psi_2}$&$3.82$~\cite{ParticleDataGroup:2024cfk}&$f_{\psi_{2}}$&$29.29^{+4.00}_{-3.69}$ GeV$^4$~\cite{Xin:2024xbp} \\
			$m_{\eta_{c2}}$&$3.83$~\cite{Xin:2024xbp}&$f_{\eta_{c2}}$&$11.64^{+1.72}_{-1.58}$ GeV$^4$~\cite{Xin:2024xbp} \\
			$m_{\psi_3}$&$3.84$~\cite{ParticleDataGroup:2024cfk}&$f_{\psi_3}$&$14.94^{+2.11}_{-1.94}$ GeV$^4$~\cite{Xin:2024xbp}\\
			$m_e$&$0.511\times10^{-3}$~\cite{ParticleDataGroup:2024cfk}&$\langle g_{s}^{2}GG\rangle$&$0.47\pm0.15$ GeV$^4$~\cite{Shifman:1978bx,Shifman:1978by}\\
			$m_\mu$&$105.7\times10^{-3}$~\cite{ParticleDataGroup:2024cfk}&$V_{cb}$&$0.041$~\cite{ParticleDataGroup:2024cfk} \\
			$m_\tau$&$1.78$~\cite{ParticleDataGroup:2024cfk}&$~$&$~$ \\
		\end{tabular}
	\end{ruledtabular}
\end{table}

From Eq.~(\ref{eq:20}), the form factor depend on the threshold parameters $s_0$ and $u_0$, Borel parameter $T^2$ and the squared of transition momentum $Q^2$. The values of $s_0$ and $u_0$ are determined by the calculations of two-point QCD sum rules which are $\sqrt{s_0}=6.56-6.71$ GeV and $\sqrt{u_0}=4.30\pm0.10$ ($4.40\pm0.10$) GeV for $\psi_1$ ($\psi_2$, $\eta_{c2}$ and $\psi_3$), respectively~\cite{Xin:2024xbp,Wang:2024fwc}. As for the Borel parameter $T^2$, an appropriate work region which is commonly called as 'Borel platform' should be chosen to obtain the final results. In this region, the dependency of the form factors on the Borel parameter should be as weak as possible. Furthermore, two conditions should be satisfied which are the pole dominance and the convergence of OPE. To define the pole contribution, we firstly write down:
\begin{eqnarray}
	\notag
	\Pi^{\mathrm{QCD}}_{\mathrm{pole}}(T^2)&&=\int\limits_{s_{\min}}^{s_0}\int\limits_{u_{\min}}^{u_0}dsdu\rho^{\mathrm{QCD}}(s,u,Q^2)\exp\left(-\frac{s}{T^2}-\frac{u}{kT^2}\right), \\
	\notag
	\Pi^{\mathrm{QCD}}_{\mathrm{cont.}}(T^2)&&=\int\limits_{s_{0}}^{\infty}\int\limits_{u_{0}}^{\infty}dsdu\rho^{\mathrm{QCD}}(s,u,Q^2)\exp\left(-\frac{s}{T^2}-\frac{u}{kT^2}\right).\\
\end{eqnarray}
Then, the pole and continuum contributions can be defined as~\cite{Bracco:2011pg}:
\begin{eqnarray}
	\notag
	\mathrm{Pole}&&=\frac{\Pi^{\mathrm{QCD}}_{\mathrm{pole}}(T^2)}{\Pi^{\mathrm{QCD}}_{\mathrm{pole}}(T^2)+\Pi^{\mathrm{QCD}}_{\mathrm{cont.}}(T^2)},\\
	\mathrm{Continuum}&&=\frac{\Pi^{\mathrm{QCD}}_{\mathrm{cont.}}(T^2)}{\Pi^{\mathrm{QCD}}_{\mathrm{pole}}(T^2)+\Pi^{\mathrm{QCD}}_{\mathrm{cont.}}(T^2)}.
\end{eqnarray}
The pole dominance requires that the pole contribution should be larger than 40$\%$ in the present work.

\begin{figure}
	\centering
	\includegraphics[width=8.5cm]{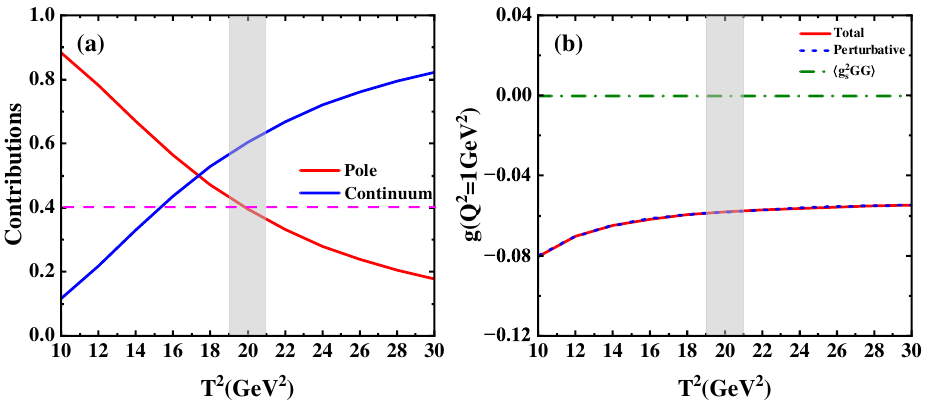}
	\caption{The pole and continuum contributions (a) and the contributions of perturbative part and gluon condensates (b) of form factor $g$ for $B_c\to\psi_1$ transition. The grey bounds are the Borel platform.}
	\label{PCandOPEC}
\end{figure}

Taking the vector form factor $g$ of $B_c\to\psi_1$ as an example, we simply discuss how to determined the Borel platform. Fixing $Q^2=1$ GeV$^2$ in Eq.~(\ref{eq:20}), we firstly plot the pole and continuum contributions of form
factor $g$ on Borel parameter $T^2$ which are shown in Fig.~\ref{PCandOPEC} (a). In addition, the contributions of perturbative part and gluon condensate for $g$ on Borel parameter $T^2$ are shown in Fig.~\ref{PCandOPEC} (b). It can be find that the pole contribution decreases as the Borel parameter increases, and Borel parameter dependency of form factor is also decreases as $T^2$ increases. Finally, the Borel platform of form factor $g$ is determined as $19-21$ GeV$^2$. The
pole contribution in Borel platform is about $40\%$. In addition, the contribution of gluon condensate term is less than $1\%$, this indicates the OPE convergence is also satisfied. After repeated trial and contrast, we finally determine the Borel platforms for all form factors in $Q^2=1$ GeV$^2$. They are $19-21$ GeV$^2$, $20-22$ GeV$^2$ and $21-23$ GeV$^2$ for the form factors related to the $B_c\to\psi_1$, $B_c\to\psi_2[\eta_{c2}]$ and $B_c\to\psi_3$ transitions, respectively.
 
\begin{figure*}
	\centering
	\includegraphics[width=18cm]{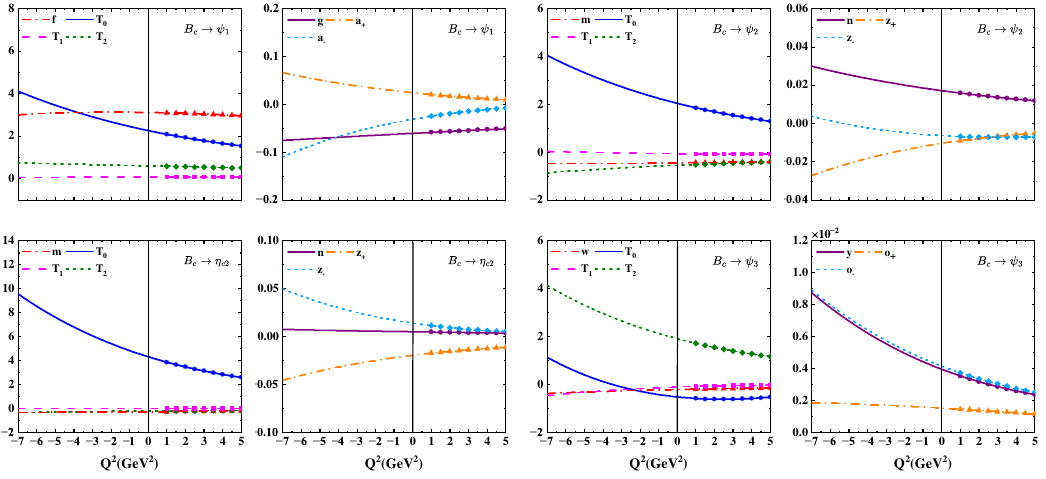}
	\caption{Fitting results of all form factors.}
	\label{FormFactors}
\end{figure*}

By taking different values of $Q^2$, the form factors in space-like regions ($Q^2>0$) will be obtained, where the range of $Q^2$ is taken as $1-5$ GeV$^2$ in present work. Then, we can use appropriate fitting models to extrapolate the form factors into time-like regions. The $z$-series expand approach~\cite{Boyd:1994tt} is introduced to carry out this work. With this approach, the form factors can be expanded as the following form,
\begin{eqnarray}
\notag
F(Q^2) &&= \frac{1}{1 + Q^2/m_{\mathrm{pole}}^2} \\
&&\times \sum\limits_{k = 0}^{N - 1} {b_k\left[ z(Q^2,t_0)^k - {(- 1)^{k - N}}\frac{k}{N} \right.} \left. z(Q^2,t_0)^N \right],
\end{eqnarray}
where $b_k$ is the fitting parameter, $m_{\mathrm{pole}}$ is taken as 6.75 GeV~\cite{Leljak:2019eyw} and $z(Q^2,t_0)$ has the following form,
\begin{eqnarray}\label{eq:28}
	z(Q^2,t_0) &&= \frac{\sqrt {t_ + + Q^2}  - \sqrt {t_ +  - t_0} }{\sqrt {t_ +  + Q^2}  + \sqrt {t_ +  - t_0}},
\end{eqnarray}
here, $t_{\pm}=(m_{B_c}\pm m_X)^2$, and $t_0=t_+-\sqrt{t_+(t_+-t_-)}$. The $z$ series are truncated at $N=3$ in present work. The central values of fitting parameters $b_k$ for all form factors are collected in Table~\ref{ZP} and the fitting results are shown in Fig.~\ref{FormFactors}. In addition, the values of all form factors in $Q^2=0$ are also presented in the last column of Table~\ref{ZP}.   The upper and lower bounds of $F(0)$ in Table~\ref{ZP} represent the combination of uncertainties arising from including threshold and Borel parameters, decay constants of mesons, and so on, with the threshold parameters and decay constants making the most significant contribution.
\begin{table}[htbp]
	\renewcommand\arraystretch{1.2}
	\begin{ruledtabular}\caption{The central values of fitting parameters in $z$ series expand approach and the values of different form factors in $Q^2=0$. The upper and lower bounds of $F(0)$ represent the combination of uncertainties arising from including threshold and Borel parameters, decay constants of mesons, and so on, with the threshold parameters and decay constants making the most significant contribution.}
		\label{ZP}
		\begin{tabular}{c c c c c c}
			Mode&FF&$b_{0}$&$b_{1}$&$b_{2}$&$F(0)$ \\ \hline
			\multirow{7}*{$B_c \to \psi_1$}&$g$ &$-0.062$&$0.19$&$5.35$&$-0.060^{+0.009}_{-0.010}$ \\
			~&$f$ &$2.92$&$31.00$&$-589.80$&$3.13^{+0.58}_{-0.49}$ \\
			~&$a_+$ &$0.037$&$-1.73$&$21.78$&$0.025^{+0.000}_{-0.001}$ \\
			~&$a_-$ &$-0.054$&$3.35$&$-52.82$&$-0.030^{+0.003}_{-0.005}$ \\
			~&$T_0$ &$2.76$&$-66.32$&$733.27$&$2.28^{+0.15}_{-0.17}$ \\
			~&$T_1$ &$0.079$&$2.60$&$-40.91$&$0.097^{+0.015}_{-0.013}$ \\
			~&$T_2$ &$0.6$&$-1.62$&$-51.82$&$0.62^{+0.11}_{-0.10}$ \\ \hline
			\multirow{7}*{$B_c \to \psi_2$}&$m$ &$-0.43$&$-2.44$&$102.41$&$-0.44^{+0.06}_{-0.07}$ \\
			~&$z_+$ &$-0.015$&$0.68$&$-10.70$&$-0.010^{+0.001}_{-0.001}$ \\
			~&$z_-$ &$-0.32\times10^{-2}$&$-0.50$&$13.95$&$(-0.63^{+0.10}_{-0.14})\times10^{-2}$ \\
			~&$n$ &$0.020$&$-0.45$&$4.25$&$0.017^{+0.002}_{-0.002}$ \\
			~&$T_0$ &$2.57$&$-75.21$&$889.27$&$2.05^{+0.21}_{-0.21}$ \\
			~&$T_1$ &$-0.019$&$-4.33$&$92.54$&$-0.046^{+0.018}_{-0.023}$ \\
			~&$T_2$ &$-0.61$&$10.03$&$-78.23$&$-0.54^{+0.05}_{-0.05}$ \\\hline
			\multirow{7}*{$B_c \to \eta_{c2}$}&$m$ &$-0.28$&$1.13$&$44.01$&$-0.27^{+0.05}_{-0.07}$ \\
			~&$z_+$ &$-0.026$&$1.04$&$-15.95$&$-0.020^{+0.002}_{-0.001}$ \\
			~&$z_-$ &$0.024$&$-1.51$&$30.34$&$0.014^{+0.001}_{-0.000}$ \\
			~&$n$ &$0.56\times10^{-2}$&$-0.07$&$-0.13$&$(0.50^{+0.11}_{-0.09})\times10^{-2}$ \\
			~&$T_0$ &$5.68$&$-204.32$&$3051.17$&$4.31^{+0.32}_{-0.36}$ \\
			~&$T_1$ &$-0.63\times10^{-3}$&$0.85\times10^{-2}$&$0.02$&$(-0.56^{+0.05}_{-0.06})\times10^{-3}$ \\
			~&$T_2$ &$-0.25$&$3.49$&$3.41$&$-0.23^{+0.04}_{-0.05}$ \\\hline
			\multirow{7}*{$B_c \to \psi_3$}&$y$ &$0.52\times10^{-2}$&$-0.19$&$2.81$&$(0.39^{+0.03}_{-0.03})\times10^{-2}$ \\
			~&$w$ &$-0.25$&$5.82$&$-62.24$&$-0.21^{+0.02}_{-0.02}$ \\
			~&$o_+$ &$0.16\times10^{-2}$&$-0.50\times10^{-2}$&$-0.52$&$(0.15^{+0.01}_{-0.01})\times10^{-2}$ \\
			~&$o_-$ &$0.54\times10^{-2}$&$-0.18$&$2.58$&$(0.41^{+0.00}_{-0.00})\times10^{-2}$ \\
			~&$T_0$ &$-0.097$&$-76.41$&$2601.08$&$-0.52^{+0.13}_{-0.17}$ \\
			~&$T_1$ &$-0.20$&$15.53$&$-324.08$&$-0.11^{+0.01}_{-0.01}$ \\
			~&$T_2$ &$2.48$&$-86.17$&$1281.06$&$1.91^{+0.15}_{-0.17}$ \\
		\end{tabular}
	\end{ruledtabular}
\end{table}

Based on above analysis of vector and axial vector form factors, we can discuss the decay widths and branching ratios of three-body semileptonic decay $B_c\to Xl\bar{\nu}_l$. The standard form of three body differential decay
width can be written as:
\begin{eqnarray}
\notag
d\Gamma(B_c\to Xl\bar{\nu}_l)&&=\frac{1}{2J+1}\sum\frac{1}{2m_{B_c}}d\Phi(p\to p',l,v)|T|^2,\\
\notag
d\Phi(p\to p',l,v)&&=(2\pi)^4\delta^4(p-p'-l-v)\\
&&\times\frac{d^3\vec{p'}}{(2\pi)^3 2p'_0}\frac{d^3\vec{l}}{(2\pi)^3 2l_0}\frac{d^3\vec{v}}{(2\pi)^3 2v_0},
\end{eqnarray}
where $J$ is the total angle momentum of $B_c$ meson, $\sum$ represents the summation of all the polarization, $d\Phi(p,p',l,v)$ denote the three-body phase space, $p$, $p'$, $l$ and $v$ are the four momentum of $B_c$, $X$, $l$ and $\bar{\nu}_l$, and $T$ is the transition matrix element. The three-body can be decomposed as the product of two-body phase space:
\begin{eqnarray}
 d\Phi(p\to p',l,v)=\frac{dq^2}{2\pi}d\Phi(p\to p',q)d\Phi(q\to l,v),
\end{eqnarray}
here, $q=l+v$ denote the four momentum of $W$ bosen, the two-body phase space can be written as:
\begin{eqnarray}
	d\Phi(p\to p',q)=(2\pi)^4\delta^4(p-p'-q)\frac{d^3\vec{p'}}{(2\pi)^3 2p'_0}\frac{d^3\vec{q}}{(2\pi)^3 2q_0}.
\end{eqnarray}
For the convenience of calculation, we introduce the helicity amplitudes which can be expressed as the following forms for the decay process $B_c\to\psi_1l\bar{\nu}_l$~\cite{Chen:2017vgi}:
\begin{eqnarray}
\notag
H^{\pm}_{\pm}(q^2)&&=if(q^2)\mp g(q^2)\sqrt{\lambda(m_{B_c}^2,m_{\psi_1}^2,q^2)},\\
\notag
H^0_0(q^2)&&=-\frac{i}{\sqrt{q^2}}\left\{{\frac{m_{B_c}^2-m_{\psi_1}^2-q^2}{2m_{\psi_1}}f(q^2)}\right.\\
\notag
&&+\left. {\frac{\lambda(m_{B_c}^2,m_{\psi_1}^2,q^2)}{2m_{\psi_1}}a_+(q^2)} \right\},\\
\notag
H^0_s(q^2)&&=-\frac{i}{\sqrt{q^2}}\frac{\sqrt{\lambda(m_{B_c}^2,m_{\psi_1}^2,q^2)}}{2m_{\psi_1}}\left[{f(q^2)} \right.\\
&&\left. {+(m_{B_c}^2-m_{\psi_1}^2)a_+(q^2) +q^2a_-(q^2)}\right].
\end{eqnarray}
where $\lambda(a,b,c)=a^2+b^2+c^2-2ab-2ac-2bc$ is the triangle function.

\begin{figure}
	\centering
	\includegraphics[width=8.5cm]{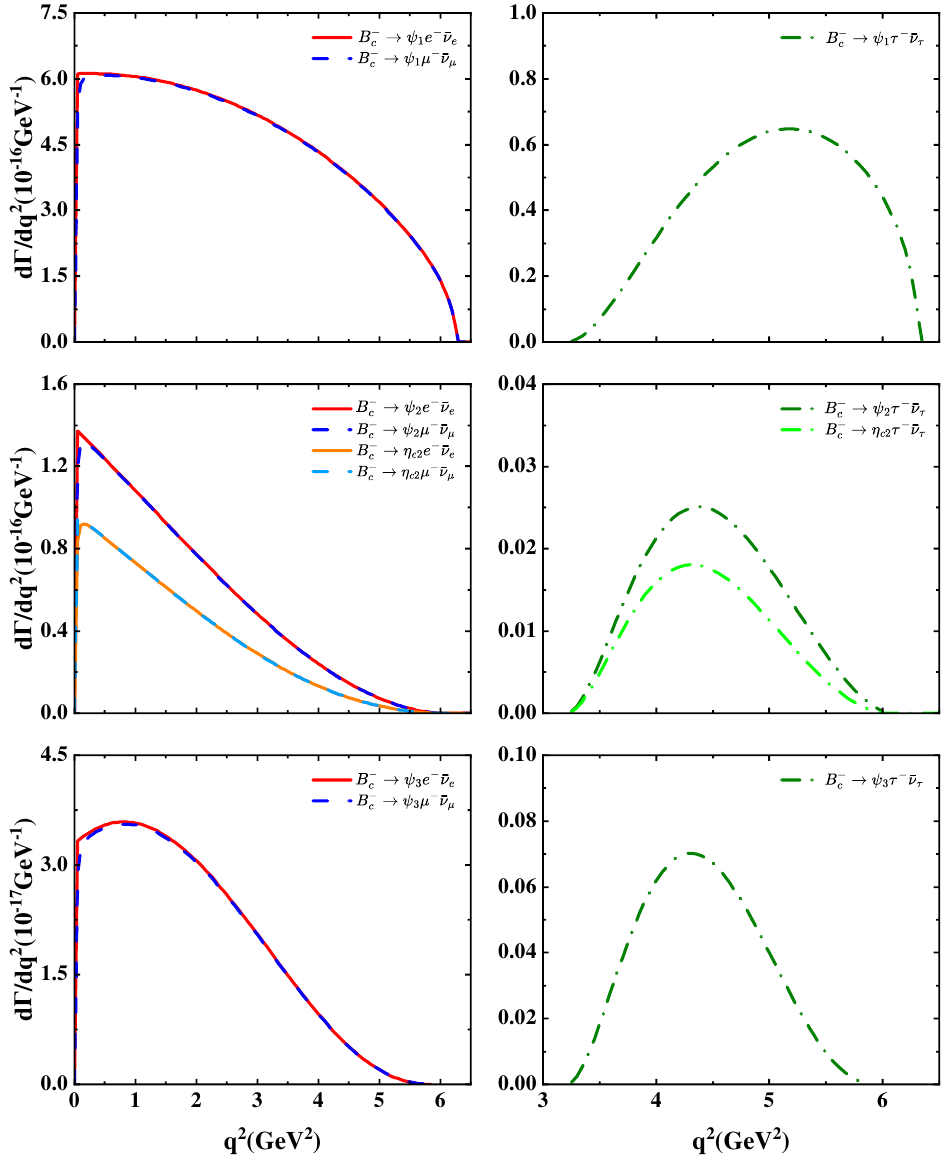}
	\caption{The differential decay width $d\Gamma/dq^2$ with variations of $q^2$ for semileptonic decay processes $B_c\to\psi_{J}[\eta_{c2}]l\bar{\nu}_l$.}
	\label{DW}
\end{figure}

Then, the differential decay width of $B_c\to\psi_1l\bar{\nu}_l$ can be expressed as the following forms:
\begin{eqnarray}
\notag
\frac{d\Gamma(B_c\to\psi_1l\bar{\nu}_l)}{dq^2}&&=\frac{d\Gamma_L(B_c\to\psi_1l\bar{\nu}_l)}{dq^2}+\frac{d\Gamma_+(B_c\to\psi_1l\bar{\nu}_l)}{dq^2}\\
\notag
&&+\frac{d\Gamma_-(B_c\to\psi_1l\bar{\nu}_l)}{dq^2},\\
\notag
\frac{d\Gamma_L(B_c\to\psi_1l\bar{\nu}_l)}{dq^2}&&=\frac{\sqrt{\lambda(m_{B_c}^2,m_{\psi_1}^2,q^2)}G_F^2V_{cb}^2}{384\pi^3m_{B_c}^3}\left(\frac{q^2-m_l^2}{q^2}\right)^2\\\notag
&&\times\left[ 3m_l^2|H_s^0|^2 + (m_l^2+2q^2)|H_0^0|^2 \right],\\
\notag
\frac{d\Gamma_\pm(B_c\to\psi_1l\bar{\nu}_l)}{dq^2}&&=\frac{\sqrt{\lambda(m_{B_c}^2,m_{\psi_1}^2,q^2)}G_F^2V_{cb}^2}{384\pi^3m_{B_c}^3}\left(\frac{q^2-m_l^2}{q^2}\right)^2\\
&&\times(m_l^2+2q^2)|H^\pm_\pm|^2,
\end{eqnarray}
where $\Gamma_L$ and $\Gamma_\pm$ denote the longitudinally and transversely polarized decay widths, respectively. For the decay processes $B_c\to\psi_2[\eta_{c2}]l\bar{\nu}_l$ and $B_c\to\psi_3l\bar{\nu}_l$, the longitudinally and transversely polarized decay widths can be obtained through the following substitutions:
\begin{eqnarray}\label{eq:34}
\notag
\frac{d\Gamma_L(B_c\to\psi_2[\eta_{c2}]l\bar{\nu}_l)}{dq^2}&&=\frac{2}{3}\frac{\lambda(m_{B_c}^2,m_{\psi_2[\eta_{c2}]}^2,q^2)}{4m_{\psi_2[\eta_{c2}]}^2}\\
\notag
&&\times\frac{d\Gamma_L(B_c\to\psi_1l\bar{\nu}_l)}{dq^2}\Bigg{|}^{g,f,a_+,a_-}_{n,m,z_+,z_-},\\
\notag
\frac{d\Gamma_\pm(B_c\to\psi_2[\eta_{c2}]l\bar{\nu}_l)}{dq^2}&&=\frac{1}{2}\frac{\lambda(m_{B_c}^2,m_{\psi_2[\eta_{c2}]}^2,q^2)}{4m_{\psi_2[\eta_{c2}]}^2}\\
\notag
&&\times\frac{d\Gamma_\pm(B_c\to\psi_1l\bar{\nu}_l)}{dq^2}\Bigg{|}^{g,f,a_+,a_-}_{n,m,z_+,z_-},
\end{eqnarray}
\begin{eqnarray}
\notag
\frac{d\Gamma_L(B_c\to\psi_3l\bar{\nu}_l)}{dq^2}&&=\frac{1}{15}\frac{\lambda(m_{B_c}^2,m_{\psi_3}^2,q^2)^2}{4m_{\psi_3}^4}\\
\notag
&&\times\frac{d\Gamma_L(B_c\to\psi_1l\bar{\nu}_l)}{dq^2}\Bigg{|}^{g,f,a_+,a_-}_{y,w,o_+,o_-},\\
\notag
\frac{d\Gamma_\pm(B_c\to\psi_3l\bar{\nu}_l)}{dq^2}&&=\frac{1}{10}\frac{\lambda(m_{B_c}^2,m_{\psi_3}^2,q^2)^2}{4m_{\psi_3}^4}\\
&&\times\frac{d\Gamma_\pm(B_c\to\psi_1l\bar{\nu}_l)}{dq^2}\Bigg{|}^{g,f,a_+,a_-}_{y,w,o_+,o_-},
\end{eqnarray}
where the superscripts and subscripts in the Eq.~(\ref{eq:34}) indicate that the form factors in the superscript should be correspondingly replaced by these in the subscript in the calculation. The differential decay widths with variations of $q^2$ for all decay processes are shown in Fig.~\ref{DW}. After finishing the integration of $q^2$, the decay width can be obtained as
\begin{eqnarray}
\Gamma=\int\limits_{m_l^2}^{(m_{B_c}-m_{\psi_J[\eta_{c2}]})^2}\frac{d\Gamma}{dq^2}dq^2.
\end{eqnarray}
The numerical results for decay widths and branching ratios are exactly shown in Table~\ref{WB}, where the uncertainties of the decay widths and branching ratios mainly origin from the form factors. From Table~\ref{WB}, We can see that $\Gamma(B_c\to \psi_1l\bar{\nu}_l)>\Gamma(B_c\to \psi_2l\bar{\nu}_l)>\Gamma(B_c\to \eta_{c2}l\bar{\nu}_l)>\Gamma(B_c\to \psi_3l\bar{\nu}_l)$, the semileptonic decay widths of $B_c$ to $D$-wave charmonia decreases as the total angular momentum of the final state $D$-wave charmonia increases. These predictions can provide valuable reference for studying the $B_c$ meson decays in future experiments by LHCb collaboration. In addition, some related decay ratios are also studied. Firstly, the central values of leptonic universality ratios are estimated, which are $R_{\psi_1}=0.045$, $R_{\psi_2}=0.012$, $R_{\eta_{c2}}=0.0091$ and $R_{\psi_3}=0.0086$, where $R_{\psi_1}$ denotes $\frac{\Gamma(B^-_c\to\psi_1\tau^-\bar{\nu}_\tau)}{\Gamma(B^-_c\to\psi_1\mu^-\bar{\nu}_\mu)}$. Then, we also calculate the central values of decay ratios of different final state mesons. For $e$ and $\mu$ decay channels, these values are $R_{\psi_1}^{\psi_2}(e/\mu)=0.12/0.11$, $R_{\psi_1}^{\psi_3}(e/\mu)=0.041/0.040$ and $R_{\psi_2}^{\eta_{c2}}(e/\mu)=0.88/0.92$, where $R_{\psi_1}^{\psi_2}(e)$ denotes $\frac{\Gamma(B^-_c\to\psi_2e^-\bar{\nu}_e)}{\Gamma(B^-_c\to\psi_1e^-\bar{\nu}_e)}$. For $\tau$ decay channel, these values are $R_{\psi_1}^{\psi_2}(\tau)=0.03$, $R_{\psi_1}^{\psi_3}(\tau)=0.0076$ and $R_{\psi_2}^{\eta_{c2}}(\tau)=0.70$. These results can provide useful information to understand the properties of $B_c$ meson and $D$-wave charmonia and to test SM.
\begin{table*}[htbp]
	\renewcommand\arraystretch{1.3}
	\begin{ruledtabular}\caption{Decay widths (in $10^{-7}$ eV) and branching ratios ($10^{-3}$) of $B_c$ to $D$-wave charmonia semileptonic decays. Branching ratios are determined by $\tau_{B_c}=0.51$ ps~\cite{ParticleDataGroup:2024cfk}.}
		\label{WB}
		\begin{tabular}{c c c c c c}
			Decay channels&Decay widths &Branching ratios&Decay channels&Decay widths &Branching ratios\\ \hline
			$B_c^-\to\psi_1e^-\bar{\nu}_e$&$28.70^{+10.70}_{-7.90}$&$2.22^{+0.84}_{-0.61}$&$B_c^-\to\eta_{c2}e^-\bar{\nu}_e$&$3.02^{+0.96}_{-0.75}$&$0.23^{+0.08}_{-0.05}$ \\
			$B_c^-\to\psi_1\mu^-\bar{\nu}_\mu$&$28.41^{+10.60}_{-7.82}$&$2.20^{+0.83}_{-0.60}$&$B_c^-\to\eta_{c2}\mu^-\bar{\nu}_\mu$&$2.97^{+0.95}_{-0.74}$&$0.23^{+0.07}_{-0.06}$ \\
			$B_c^-\to\psi_1\tau^-\bar{\nu}_\tau$&$1.31^{+0.54}_{-0.38}$&$0.10^{+0.04}_{-0.03}$&	$B_c^-\to\eta_{c2}\tau^-\bar{\nu}_\tau$&$0.027^{+0.011}_{-0.007}$&$0.21^{+0.08}_{-0.06}\times10^{-2}$ \\
			$B_c^-\to\psi_2e^-\bar{\nu}_e$&$3.32^{+0.89}_{-0.85}$&$0.26^{+0.07}_{-0.07}$&$B_c^-\to\psi_3e^-\bar{\nu}_e$&$1.15^{+0.31}_{-0.28}$&$0.090^{+0.020}_{-0.022}$ \\
			$B_c^-\to\psi_2\mu^-\bar{\nu}_\mu$&$3.27^{+0.88}_{-0.84}$&$0.25^{+0.07}_{-0.06}$&$B_c^-\to\psi_3\mu^-\bar{\nu}_\mu$&$1.14^{+0.30}_{-0.28}$&$0.088^{+0.022}_{-0.021}$ \\
			$B_c^-\to\psi_2\tau^-\bar{\nu}_\tau$&$0.039^{+0.012}_{-0.010}$&$0.30^{+0.09}_{-0.08}\times10^{-2}$&$B_c^-\to\psi_3\tau^-\bar{\nu}_\tau$&$0.98^{+0.26}_{-0.22}\times10^{-2}$&$0.76^{+0.20}_{-0.17}\times10^{-3}$ \\
		\end{tabular}
	\end{ruledtabular}
\end{table*}

\section{conclusion}\label{sec5}
In this work, we analyze the vector, axial vector and tensor form factors of $B_c$ to $D$-wave charmonia in space-like region using the three-point QCD sum rules. These form factors are then extrapolated to the full physical region via the 
$z$-series expansion method. With the estimated vector and axial vector form factors, we perform the numerical calculation of decay widths and branching ratios for the semileptonic decays of $B_c$ to $D$-wave charmonia. These predictions can provide valuable references for future studies of $B_c$ meson at LHCb experiments. It is also expected that these results can help to shed more light on the properties of $B_c$ meson and $D$-wave charmonia, thereby providing useful information to research
the heavy-flavor physics.

\section*{Acknowledgments}
This work is supported by National Natural Science Foundation of China under the Grant No. 12575083, as well as supported, in part, by National Key Research and Development Program under Grant No. 2024YFA1610503 and Natural Science Foundation of HeBei Province under the Grant No. A2024502002.

\begin{widetext}
\appendix
\section{The expressions of sum rules for form factors.}\label{Sec:AppA}
The full expressions of sum rules for form factors are as follows:
\begin{eqnarray}
\notag
f^{B_c \to \psi _1}(Q^2) &&=\frac{m_b + m_c}{f_{\psi _1}f_{B_c}m_{B_c}^2}\exp \left( \frac{m_{B_c}^2}{T^2} +\frac{m_{\psi _1}^2}{kT^2} \right)\int\limits_{s_{\min }}^{s_0} \int\limits_{u_{\min }}^{u_0} dsdu  \rho _2^{\psi _1\mathrm{QCD}}(s,u,Q^2)\exp \left( - \frac{s}{T^2} - \frac{u}{kT^2} \right),\\
\notag
a_ + ^{B_c \to \psi _1}(Q^2) &&=\frac{m_b + m_c}{2f_{\psi _1}f_{B_c}m_{B_c}^2}\exp \left( \frac{m_{B_c}^2}{T^2} + \frac{m_{\psi _1}^2}{kT^2} \right)\int\limits_{s_{\min }}^{s_0} \int\limits_{u_{\min }}^{u_0} dsdu \left[ \rho _3^{\psi _1\mathrm{QCD}}(s,u,Q^2) + \rho _4^{\psi _1\mathrm{QCD}}(s,u,Q^2) \right]\exp \left(  - \frac{s}{T^2} - \frac{u}{kT^2} \right),\\
a_ - ^{B_c \to \psi _1}(Q^2) &&=\frac{m_b + m_c}{2f_{\psi _1}f_{B_c}m_{B_c}^2}\exp \left( \frac{m_{B_c}^2}{T^2} + \frac{m_{\psi _1}^2}{kT^2} \right)\int\limits_{s_{\min }}^{s_0} \int\limits_{u_{\min }}^{u_0} dsdu \left[ \rho _3^{\psi _1\mathrm{QCD}}(s,u,Q^2) - \rho _4^{\psi _1\mathrm{QCD}}(s,u,Q^2) \right]\exp \left( - \frac{s}{T^2} - \frac{u}{kT^2} \right),
\end{eqnarray}
\begin{eqnarray}
\notag
T_0^{B_c \to \psi _1}(Q^2) &&=-\frac{(m_b + m_c)(m_{B_c} + m_{\psi _1})^2}{f_{\psi _1}f_{B_c}m_{B_c}^2}\exp \left( \frac{m_{B_c}^2}{T^2} + \frac{m_{\psi _1}^2}{kT^2} \right)\int\limits_{s_{\min }}^{s_0} \int\limits_{u_{\min }}^{u_0} dsdu \rho _5^{\psi _1\mathrm{QCD}}(s,u,Q^2)\exp \left( - \frac{s}{T^2} - \frac{u}{kT^2} \right),\\
\notag
T_1^{B_c \to \psi _1}(Q^2) &&=-\frac{m_b + m_c}{f_{\psi _1}f_{B_c}m_{B_c}^2}\exp \left( \frac{m_{B_c}^2}{T^2} + \frac{m_{\psi _1}^2}{kT^2} \right)\int\limits_{s_{\min }}^{s_0} \int\limits_{u_{\min }}^{u_0} dsdu \rho _6^{\psi _1\mathrm{QCD}}(s,u,Q^2)\exp \left( - \frac{s}{T^2} - \frac{u}{kT^2} \right),\\
T_2^{B_c \to \psi _1}(Q^2) &&=-\frac{m_b + m_c}{f_{\psi _1}f_{B_c}m_{B_c}^2}\exp \left( \frac{m_{B_c}^2}{T^2} + \frac{m_{\psi _1}^2}{kT^2} \right)\int\limits_{s_{\min }}^{s_0} \int\limits_{u_{\min }}^{u_0} dsdu \rho _7^{\psi _1\mathrm{QCD}}(s,u,Q^2)\exp \left( - \frac{s}{T^2} - \frac{u}{kT^2} \right),
\end{eqnarray}
\begin{eqnarray}
\notag
m^{B_c \to \psi _2[\eta _{c2}]}(Q^2) &&= \frac{2(m_b + m_c)}{f_{\psi _2[\eta _{c2}]}f_{B_c}m_{B_c}^2}\exp \left( \frac{m_{B_c}^2}{T^2} + \frac{m_{\psi _2[\eta _{c2}]}^2}{kT^2}\right)\int\limits_{s_{\min}}^{s_0} \int\limits_{u_{\min }}^{u_0} dsdu \rho _1^{\psi _2[\eta _{c2}]\mathrm{QCD}}(s,u,Q^2)\exp \left( - \frac{s}{T^2} - \frac{u}{kT^2} \right),\\
\notag
z_ + ^{B_c \to \psi _2[\eta _{c2}]}(Q^2) &&= \frac{m_b + m_c}{2f_{\psi _2[\eta _{c2}]}f_{B_c}m_{B_c}^2}\exp \left( \frac{m_{B_c}^2}{T^2} + \frac{m_{\psi _2[\eta _{c2}]}^2}{kT^2} \right)\\
\notag
&&\times\int\limits_{s_{\min }}^{s_0} \int\limits_{u_{\min }}^{u_0} dsdu \left[ \rho _2^{\psi _2[\eta _{c2}]\mathrm{QCD}}(s,u,Q^2) + \rho _3^{\psi _2[\eta _{c2}]\mathrm{QCD}}(s,u,Q^2) \right]\exp \left( - \frac{s}{T^2} - \frac{u}{kT^2} \right),\\
\notag
z_ - ^{B_c \to \psi _2[\eta _{c2}]}(Q^2) &&= \frac{m_b + m_c}{2f_{\psi _2[\eta _{c2}]}f_{B_c}m_{B_c}^2}\exp \left( \frac{m_{B_c}^2}{T^2} + \frac{m_{\psi _2[\eta _{c2}]}^2}{kT^2} \right)\\
&&\times\int\limits_{s_{\min }}^{s_0} \int\limits_{u_{\min }}^{u_0} dsdu \left[ \rho _2^{\psi _2[\eta _{c2}]\mathrm{QCD}}(s,u,Q^2) - \rho _3^{\psi _2[\eta _{c2}]\mathrm{QCD}}(s,u,Q^2) \right]\exp \left( - \frac{s}{T^2} - \frac{u}{kT^2} \right),
\end{eqnarray}
\begin{eqnarray}
n^{B_c \to \psi _2[\eta _{c2}]}(Q^2) =-\frac{m_b + m_c}{f_{\psi _2[\eta _{c2}]}f_{B_c}m_{B_c}^2}\exp \left( \frac{m_{B_c}^2}{T^2} + \frac{m_{\psi _2[\eta _{c2}]}^2}{kT^2} \right)\int\limits_{s_{\min }}^{s_0} \int\limits_{u_{\min }}^{u_0} dsdu \rho _4^{\psi _2[\eta _{c2}]\mathrm{QCD}}(s,u,Q^2)\exp \left( - \frac{s}{T^2} - \frac{u}{kT^2} \right),
\end{eqnarray}
\begin{eqnarray}
\notag
T_0^{B_c \to \psi _2[\eta _{c2}]}(Q^2) &&=-\frac{(m_b + m_c){(m_{B_c} + m_{\psi _2[\eta _{c2}]})}^2}{f_{\psi _2[\eta _{c2}]}f_{B_c}m_{B_c}}\exp \left( \frac{m_{B_c}^2}{T^2} + \frac{m_{\psi _2[\eta _{c2}]}^2}{kT^2} \right)\int\limits_{s_{\min }}^{s_0} \int\limits_{u_{\min }}^{u_0} dsdu \rho _5^{\psi _2[\eta _{c2}]\mathrm{QCD}}(s,u,Q^2)\exp \left( - \frac{s}{T^2} - \frac{u}{kT^2} \right),\\
\notag
T_1^{B_c \to \psi _2[\eta _{c2}]}(Q^2) &&=-\frac{2(m_b + m_c)}{f_{\psi _2[\eta _{c2}]}f_{B_c}m_{B_c}}\exp \left( \frac{m_{B_c}^2}{T^2} + \frac{m_{\psi _2[\eta _{c2}]}^2}{kT^2} \right)\int\limits_{s_{\min }}^{s_0} \int\limits_{u_{\min }}^{u_0} dsdu \rho _6^{\psi _2[\eta _{c2}]\mathrm{QCD}}(s,u,Q^2)\exp \left( - \frac{s}{T^2} - \frac{u}{kT^2} \right),\\
T_2^{B_c \to \psi _2[\eta _{c2}]}(Q^2) &&=-\frac{2(m_b + m_c)}{f_{\psi _2[\eta _{c2}]}f_{B_c}m_{B_c}}\exp \left( \frac{m_{B_c}^2}{T^2} + \frac{m_{\psi _2[\eta _{c2}]}^2}{kT^2} \right)\int\limits_{s_{\min }}^{s_0} \int\limits_{u_{\min }}^{u_0} dsdu \rho _7^{\psi _2[\eta _{c2}]\mathrm{QCD}}(s,u,Q^2)\exp \left( - \frac{s}{T^2} - \frac{u}{kT^2} \right),
\end{eqnarray}
\begin{eqnarray}
y^{B_c \to \psi _3}(Q^2) = \frac{3(m_b + m_c)}{2f_{\psi _3}f_{B_c}m_{B_c}^2}\exp \left( \frac{m_{B_c}^2}{T^2} + \frac{m_{\psi _3}^2}{kT^2} \right)\int\limits_{s_{\min }}^{s_0} \int\limits_{u_{\min }}^{u_0} dsdu \rho _1^{\psi _3\mathrm{QCD}}(s,u,Q^2)\exp \left( - \frac{s}{T^2} - \frac{u}{kT^2} \right),
\end{eqnarray}
\begin{eqnarray}
\notag
w^{B_c \to \psi _3}(Q^2) &&=  \frac{3(m_b + m_c)}{f_{\psi _3}f_{B_c}m_{B_c}^2}\exp \left( \frac{m_{B_c}^2}{T^2} + \frac{m_{\psi _3}^2}{kT^2} \right)\int\limits_{s_{\min }}^{s_0} \int\limits_{u_{\min }}^{u_0} dsdu \rho _2^{\psi _3\mathrm{QCD}}(s,u,Q^2)\exp \left( - \frac{s}{T^2} - \frac{u}{kT^2} \right),\\
\notag
o_ + ^{B_c \to \psi _3}(Q^2) &&= \frac{m_b + m_c}{2f_{\psi _3}f_{B_c}m_{B_c}^2}\exp \left( \frac{m_{B_c}^2}{T^2} + \frac{m_{\psi _3}^2}{kT^2} \right)\int\limits_{s_{\min }}^{s_0} \int\limits_{u_{\min }}^{u_0} dsdu \left[ \rho _3^{\psi _3\mathrm{QCD}}(s,u,Q^2) + \rho _4^{\psi _3\mathrm{QCD}}(s,u,Q^2) \right]\exp \left(  - \frac{s}{T^2} - \frac{u}{kT^2} \right),\\
o_ - ^{B_c \to \psi _3}(Q^2) &&= \frac{m_b + m_c}{2f_{\psi _3}f_{B_c}m_{B_c}^2}\exp \left( \frac{m_{B_c}^2}{T^2} + \frac{m_{\psi _3}^2}{kT^2} \right)\int\limits_{s_{\min }}^{s_0} \int\limits_{u_{\min }}^{u_0} dsdu \left[ \rho _3^{\psi _3\mathrm{QCD}}(s,u,Q^2) - \rho _4^{\psi _3\mathrm{QCD}}(s,u,Q^2) \right]\exp \left(  - \frac{s}{T^2} - \frac{u}{kT^2} \right),
\end{eqnarray}
\begin{eqnarray}
\notag
T_0^{B_c \to \psi _3}(Q^2) &&=-\frac{(m_b + m_c)(m_{B_c} + m_{\psi _3})^2}{f_{\psi _3}f_{B_c}}\exp \left( \frac{m_{B_c}^2}{T^2} + \frac{m_{\psi _3}^2}{kT^2} \right)\int\limits_{s_{\min }}^{s_0} \int\limits_{u_{\min }}^{u_0} dsdu \rho _5^{\psi _3\mathrm{QCD}}(s,u,Q^2)\exp \left( - \frac{s}{T^2} - \frac{u}{kT^2} \right),\\
\notag
T_1^{B_c \to \psi _3}(Q^2) &&=-\frac{3(m_b + m_c)}{f_{\psi _3}f_{B_c}}\exp \left( \frac{m_{B_c}^2}{T^2} + \frac{m_{\psi _3}^2}{kT^2} \right)\int\limits_{s_{\min }}^{s_0} \int\limits_{u_{\min }}^{u_0} dsdu \rho _6^{\psi _3\mathrm{QCD}}(s,u,Q^2)\exp \left( - \frac{s}{T^2} - \frac{u}{kT^2} \right),\\
T_2^{B_c \to \psi _3}(Q^2) &&=-\frac{3(m_b +m_c)}{f_{\psi _3}f_{B_c}}\exp \left( \frac{m_{B_c}^2}{T^2} + \frac{m_{\psi _3}^2}{kT^2} \right)\int\limits_{s_{\min }}^{s_0} \int\limits_{u_{\min }}^{u_0} dsdu \rho _7^{\psi _3\mathrm{QCD}}(s,u,Q^2)\exp \left( - \frac{s}{T^2} - \frac{u}{kT^2} \right),
\end{eqnarray}
where the QCD spectral densities $\rho^{\mathrm{QCD}}_i$ in above equations are obtained by analyzing the invariant amplitudes $\Pi^{\mathrm{QCD}}_i$ in Eqs.~(\ref{eq:15})-(\ref{eq:17}).
\end{widetext}

\bibliography{ref.bib}

\end{document}